\documentclass[english,aps,prb,superscriptaddress,reprint,longbibliograhy]{revtex4-2}

\usepackage{graphicx}  
\usepackage{float}     
\usepackage{dcolumn}   
\usepackage{bm}        
\usepackage{amssymb}   
\usepackage{amsmath}
\usepackage{dsfont}
\usepackage{lipsum}    
\usepackage{xcolor}
\usepackage{soul}
\usepackage{float}
\usepackage{placeins}
\usepackage{physics}
\usepackage{mathrsfs}
\usepackage{bbold}
\usepackage{hyperref}
\hypersetup{
    colorlinks=true,     
    linkcolor=blue,      
    citecolor=blue,      
    urlcolor=blue,       
    }

\usepackage[utf8]{inputenc}
\usepackage{natbib}
\usepackage{graphicx}
\usepackage{subcaption}
\usepackage{booktabs}
\usepackage{blindtext}

\usepackage{color}

\definecolor{goodred}{RGB}{183,15,58}
\definecolor{goodblue}{RGB}{93,128,180}

\begin{document}

\title{\Large\textbf{Momentum-dependent precessional and nutational spin pumping in a honeycomb antiferromagnet}}

\author{Suman Mukherjee}
\affiliation{Department of Physics, Indian Institute of Technology (ISM) Dhanbad, IN-826004, Dhanbad, India}

\author{Subhadip Ghosh}
\affiliation{Department of Physics, Indian Institute of Technology (ISM) Dhanbad, IN-826004, Dhanbad, India}

\author{Ritwik Mondal}
\email[]{ritwik@iitism.ac.in}
\affiliation{Department of Physics, Indian Institute of Technology (ISM) Dhanbad, IN-826004, Dhanbad, India}

\begin{abstract}
The ultrafast magnetic inertial dynamics on subpicosecond timescales generate an additional high-frequency terahertz nutational resonance. Here, we investigate momentum-resolved spin pumping in a two-dimensional honeycomb antiferromagnet by incorporating spin inertia into the Landau-Lifshitz-Gilbert equation. Using a microscopic two-sublattice model with $J_1$-$J_2$-$J_3$ exchange interactions, we calculate the precessional and nutational magnon spectra and their corresponding intra- and inter-sublattice spin pumping contributions throughout the Brillouin zone. For parameters representative of MnPS$_3$, we find a pronounced complementary momentum dependence of the two spin pumping channels: the precessional current is strongly suppressed around the $\Gamma$ point ($0.01\%$) and increases towards the Brillouin zone boundary ($100\%$ at $K$ and $87\%$ at $M$), whereas the nutational current is largest near $\Gamma$ ($100\%$) and decreases towards the boundary ($68\%$ at $K$ and $69\%$ at $M$). This contrasting momentum dependence provides a means of distinguishing spin nutation from conventional precessional motion. We further demonstrate that the nutational spin pumping current is strongly controlled by the inertial relaxation time $\eta$, in contrast to the comparatively weak $\eta$ dependence of the precessional current. In the small-$\eta$ regime, our analytical results show that the intra-sublattice nutational spin pumping current exhibits a leading-order dependence of $1/\eta^3$. The spin pumping response can also be tuned through the exchange interaction strengths and magnetic moment. Our further results on the magnetic field dependence of spin pumping reveal that the ratio of nutational spin pumping current increases with magnetic field strength, whereas the corresponding precessional counterpart decreases. These results establish momentum- and field-dependent spin pumping as a promising route for identifying and controlling inertial magnon dynamics in two-dimensional antiferromagnets.

\end{abstract}
\maketitle
      
\section{Introduction}
Spin pumping has emerged as one of the most efficient mechanisms for generating spin currents by transferring angular momentum from a dynamic magnetic system into an adjacent nonmagnetic conductor~\cite{Yaroslav2002PRB,Mosendz2010,Chen2015PRL,Tserkovnyak2005,Maekawa2012SpinCurrent,Heinrich2011}. Owing to its intimate connection with magnetization dynamics, spin pumping has become a powerful probe of spin transport and interfacial spin transfer, with the resulting spin current being detected electrically through the inverse spin Hall effect~\cite{Ando2011,Sanchez2014PRL,Sanchez2013PRB,Liu2021APL,Gupta2024,Kajiwara2010}. These advances have established spin pumping as a versatile platform for investigating dynamical spin phenomena in a broad range of magnetic materials.

Among these, antiferromagnets have recently emerged as promising spin current sources owing to their vanishing net magnetization, robustness against external magnetic fields, and intrinsic spin dynamics extending into the terahertz frequency regime~\cite{Jungwirth2016,Baltz2018,Kampfrath2011,DalDin2024,Hirohata2022Advances,XIONG2022522}. Their ultrafast dynamics, together with the absence of stray fields, make antiferromagnets attractive candidates for high-density and energy-efficient spintronic devices. Beyond conventional collinear antiferromagnets, noncollinear antiferromagnets have further broadened the landscape of antiferromagnetic spintronics by exhibiting unconventional spin textures, symmetry-driven transport phenomena, and large Berry-curvature-induced responses, thereby enabling efficient spin-current generation and manipulation~\cite{Rimmler2025,Libor2020,Yang_2017}. Consequently, considerable theoretical and experimental efforts have been devoted to understanding spin transport phenomena in antiferromagnets, including spin pumping, spin-transfer torque, and spin Seebeck effects~\cite{GAN2026174373,Cheng2014PRL,Takei2014,Subedi2025,Wang2021,Vaidya2020Subterahertz,Kholid2023Interface}. In particular, it has been demonstrated that the coherent dynamics of the N\'eel order can efficiently pump pure spin currents into an adjacent normal metal despite the absence of a net equilibrium magnetization~\cite{Cheng2014PRL,Chen2021MagnonValve,Ominato_2025,Varela-Manjarres_2023,Alliati2022MnPS3,Johansen2017}. Note that recently discovered altermagnets can also inject (receive) spin currents into (from) normal metals~\cite{Hodt2024,Cho2023}.

Spin pumping is intrinsically governed by the underlying magnetization dynamics, as the time-dependent magnetic order transfers angular momentum across a magnetic interface. In conventional magnetic systems, this dynamical response is dominated by the coherent precessional motion of the magnetic moments about their equilibrium direction, giving rise to the well-established precessional spin pumping mechanism. In ferromagnets, such precessional dynamics typically occur in the gigahertz frequency regime, thereby limiting the operating speed and the rate of spin-angular-momentum transfer. The inclusion of inertial effects in the Landau--Lifshitz--Gilbert (LLG) equation, however, predicts the existence of an additional high-frequency nutational mode originating from the finite angular momentum relaxation time of the magnetic moments~\cite{Ciornei2011,Olive2015,Fahnle2011}. Unlike the conventional precessional resonance, the nutational resonance lies in the terahertz frequency regime. It constitutes a fundamentally distinct mode of magnetization dynamics~\cite{Kachkachi2025,Rodriguez2024PRL,He2023,He2024PRB,He2025,Ghosh2026PRR}. Such nutational resonance has been experimentally observed. The angular momentum relaxation time is found to be $\sim$100 fs for epitaxial Co~\cite{unikandanunni2021inertial}, $\sim$300 fs for CoFeB~\cite{Neeraj2020}, $\sim$ 1.2 ps for Permalloy~\cite{De2025PRB}. The theoretical understanding of such nutation resonance has been attributed to several microscopic mechanisms e.g., higher-order relativistic spin-orbit coupling~\cite{Mondal2017Nutation,Mondal2018}, bath-induced spin inertia~\cite{Quarenta2024}, unquenched orbital angular momentum~\cite{Moussa2026}, dynamical exchange mediated by the conduction electrons~\cite{Jansen2025,Kachkachi2025}, intrinsic non-Markovian magnetization dynamics~\cite{Hartmann2025NonMarkovian}, quantum transport formalism~\cite{Bhattacharjee_2012,Bajpai2019} and others~\cite{Kikuchi2015,thonig2017magnetic,Ghosh2024,cherkasskii2020nutation,Bajaj2024,Olive2012,Titov2024JAP,Mondal2021JPCM,Titov_2022,Thibaudeau2021,Winter2022,Cherkasskii2022Anisotropy}. In antiferromagnets, the nutational resonance is exchange-enhanced, resulting in a pronounced dissipation peak that is substantially stronger than in ferromagnets~\cite{Mondal2021PRB,Mondal2020nutation}.  

Spin pumping at the nutation resonance has been investigated in both ferromagnets and antiferromagnets~\cite{Mondal2021PRBSpinCurrent}. While a small spin pumping current is generated at the nutation resonance in ferromagnets, the corresponding current is strongly exchange enhanced in antiferromagnets~\cite{Mondal2021PRBSpinCurrent}. Moreover, the spin pumping current generated by the nutational mode is found to have an opposite sign to that associated with the conventional precessional mode. However, these studies considered spatially uniform dynamics at the $\Gamma$ point and did not incorporate the underlying crystal lattice or the momentum-dependent magnon spectrum. Consequently, how the precessional and nutational spin pumping currents evolve across the Brillouin zone remains unexplored.

In this work, we investigate the momentum-dependent precessional and nutational spin pumping currents in a two-dimensional antiferromagnet with a honeycomb crystal structure. To proceed, we consider isotropic nearest-neighbor, next-nearest-neighbor, and next-next-nearest-neighbor exchange interactions, with the equilibrium magnetization oriented out of the plane, representing the structure of MnPS$_3$~\cite{Wildes2006}. Using the linearized inertial LLG equation, we calculate the precessional and nutational magnon spectra along the high-symmetry path of the magnetic Brillouin zone, i.e., $\Gamma - K - M - \Gamma$. Further, we compute the precessional and nutational spin pumping currents and map them onto the magnon spectra to elucidate their momentum dependence across the Brillouin zone, including both intra- and inter-sublattice contributions. 

We find that the intra-sublattice spin pumping current associated with the precessional mode increases progressively from the $\Gamma$ point toward the Brillouin zone boundaries, while remaining negligibly small near $\Gamma$. In contrast, the nutational spin pumping current exhibits the opposite trend, reaching its maximum near the $\Gamma$ point and decreasing toward the Brillouin zone boundaries. Despite this decrease, the nutational current remains comparatively robust across the Brillouin zone, compared to the pronounced momentum dependence of the precessional current. The inter-sublattice contribution further distinguishes the two dynamical modes. For the precessional mode, the inter-sublattice spin pumping current has an opposite sign to the intra-sublattice contribution. It vanishes at the Brillouin zone boundaries and remains strongly suppressed near the $\Gamma$ point. In contrast, the inter-sublattice nutational spin pumping current has the same sign as its intra-sublattice counterpart. The two contributions therefore add constructively near the $\Gamma$ point, resulting in an enhanced total nutational spin pumping current. More importantly, the absence of significant cancellation between the intra- and inter-sublattice contributions allows the nutational current to remain robust throughout the Brillouin zone, whereas the precessional current becomes dominant primarily near the Brillouin zone boundaries. This distinct momentum dependence provides a clear means of distinguishing precessional and nutational spin pumping currents and may provide an experimentally accessible signature of nutational spin pumping. We further confirm these features by mapping the spin pumping currents across the two-dimensional Brillouin zone.   

We then investigate the dependence of the total spin pumping current, including both intra- and inter-sublattice contributions, on key material parameters, namely the magnetic moment, exchange interaction strengths, and inertial relaxation time. We find that varying the magnetic moment has little influence on the nutational spin pumping current at both the $\Gamma$ and $K$ points. In contrast, the precessional spin pumping current increases significantly with increasing magnetic moment at the $K$ point, while remaining strongly suppressed near the $\Gamma$ point. We next examine the dependence on the inertial relaxation time $\eta$. Increasing $\eta$ lowers the nutational frequency and consequently enhances the nutational spin pumping current at both $\Gamma$ and $K$. To further characterize this momentum-dependent response, we calculate the ratio of the nutational spin pumping currents at $\Gamma$ and $K$ for several values of the nearest-neighbor and next-nearest-neighbor exchange interactions. The ratio increases systematically with $\eta$, indicating that the nutational current at $\Gamma$ grows more rapidly with increasing inertial relaxation time than that at $K$. This distinct momentum dependence clearly separates the nutational spin pumping response from the conventional precessional contribution.


We further investigate the effect of a static magnetic field on the spin pumping current. The applied magnetic field breaks the symmetry between the two antiparallel magnetic sublattices and consequently lifts the degeneracy of the magnon bands. Interestingly, the spin pumping current is not distributed equally between the two resulting non-degenerate branches. To quantify this field-induced redistribution, we calculate the ratio of the spin pumping currents associated with the two branches. We find that the ratio of the nutational spin pumping currents increases with increasing magnetic field, whereas the corresponding ratio for the precessional spin pumping currents decreases. This contrasting field dependence provides an experimentally accessible means of distinguishing the two dynamical channels.

\section{Theory of spin pumping current }\label{Sec2}

We consider a two-dimensional honeycomb antiferromagnet described by a classical atomistic spin Hamiltonian
\begin{align}
\mathcal{H}
&=
\sum_{i\in A,j\in B}J_{ij}
\mathbf{S}_{i}^{A}\cdot\mathbf{S}_{j}^{B}+K_A\sum_{i\in A}\left(S_{i,z}^{A}\right)^2+
K_B\sum_{j\in B}\left(S_{j,z}^{B}\right)^2\nonumber\\
&
-\mu_s\sum_{i\in A,j\in B}\mathbf{B}\cdot\left[\mathbf{S}_{i}^{A} +\mathbf{S}_{j}^{B}\right],\label{Eq1}
\end{align}
where, $\mathbf{S}_{i}^{A(B)}$ denotes a unit vector representing the magnetic moment on sublattice $A(B)$, $J_{ij}$ is the exchange interaction, $K_A$ and $K_B$ are the uniaxial easy-axis anisotropies, $\mu_s$ is the magnetic moment per magnetic ion, and $\mathbf{B}$ is the external magnetic field. Throughout this work, the model parameters are chosen to represent the honeycomb antiferromagnet MnPS$_3$ as given in Table~\ref{table1}, while the theoretical framework developed here is generally applicable to other two-dimensional antiferromagnetic materials.

The honeycomb lattice in MnPS$_3$ consists of two equivalent magnetic sublattices,
denoted by $A$ and $B$, which form a bipartite structure. Each site on
the $A$ sublattice has three nearest-neighbor (NN) sites belonging to the $B$ sublattice,
while each site has six next-nearest-neighbor 
(NNN) sites belonging to the same sublattice and
three next-to-next-nearest-neighbor (NNNN) sites belonging to the opposite sublattice. Thus, the
$J_1$ and $J_3$ interactions couple the two different sublattices,
whereas the $J_2$ interaction acts within the same sublattice. The three NN vectors are   $\boldsymbol{\delta}_1 =-a\hat{\mathbf{x}},
\boldsymbol{\delta}_2=
\frac{a}{2}\hat{\mathbf{x}}
-\frac{\sqrt{3}a}{2}\hat{\mathbf{y}},
\boldsymbol{\delta}_3 =
\frac{a}{2}\hat{\mathbf{x}}
+\frac{\sqrt{3}a}{2}\hat{\mathbf{y}}$. The six NNN vectors are  $\mathbf{a}_1=
-\frac{3a}{2}\hat{\mathbf{x}}
-\frac{\sqrt{3}a}{2}\hat{\mathbf{y}},
\mathbf{a}_2=
-\sqrt{3}a\hat{\mathbf{y}},
\mathbf{a}_3=
\frac{3a}{2}\hat{\mathbf{x}}
-\frac{\sqrt{3}a}{2}\hat{\mathbf{y}},
\mathbf{a}_4=
\frac{3a}{2}\hat{\mathbf{x}}
+\frac{\sqrt{3}a}{2}\hat{\mathbf{y}},
\mathbf{a}_5=
\sqrt{3}a\hat{\mathbf{y}},
\mathbf{a}_6=
-\frac{3a}{2}\hat{\mathbf{x}}
+\frac{\sqrt{3}a}{2}\hat{\mathbf{y}}$. Finally, the three NNNN vectors are $\mathbf{b}_1=
-a\hat{\mathbf{x}}
-\sqrt{3}a\hat{\mathbf{y}},
\mathbf{b}_2=
2a\hat{\mathbf{x}},
\mathbf{b}_3=
-a\hat{\mathbf{x}}
+\sqrt{3}a\hat{\mathbf{y}}$. Here $a$ is the lattice constant. 
\begin{figure}[h!]
\includegraphics[width=0.95\linewidth]{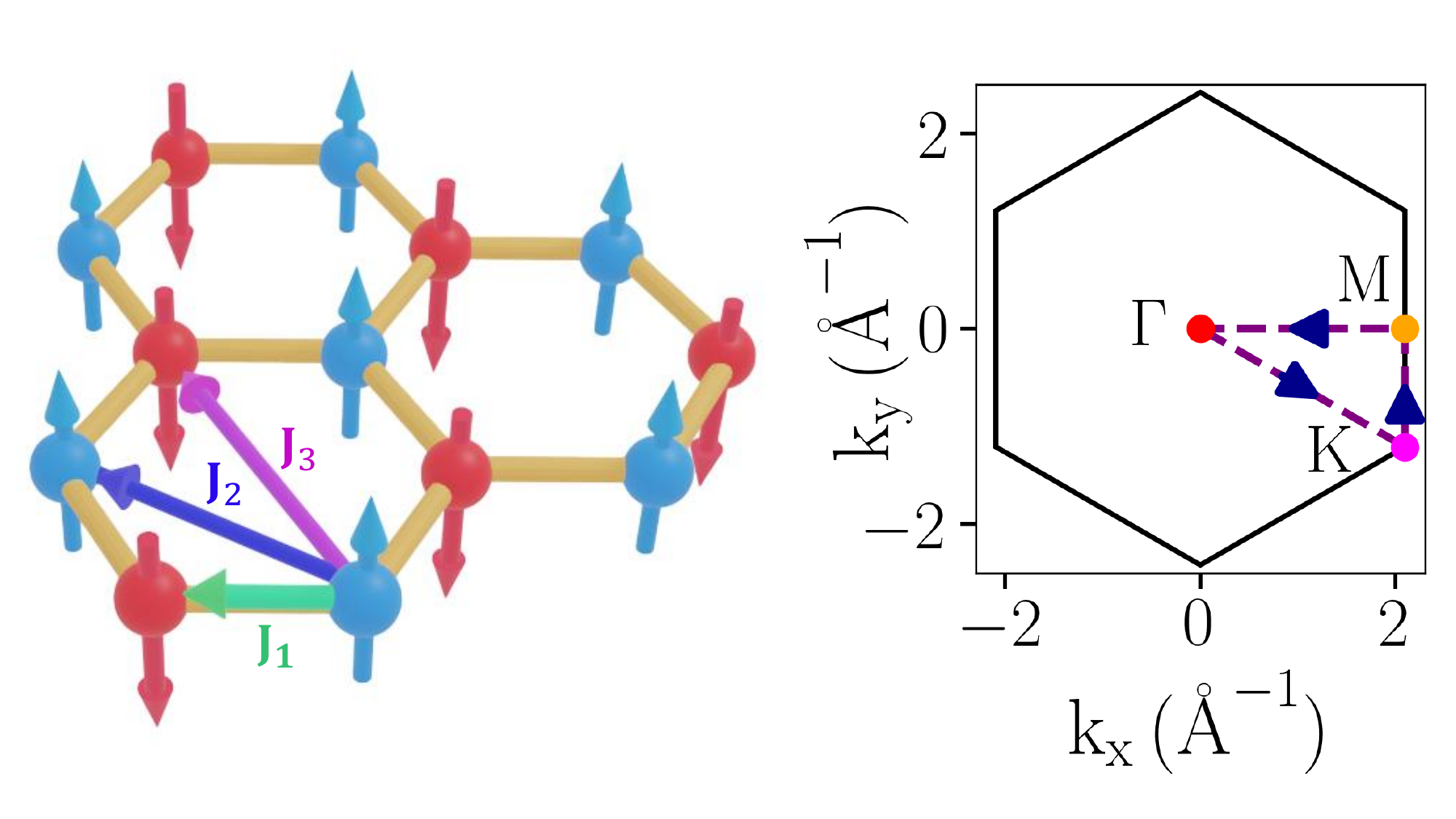}
    \caption{Schematic illustration of two-dimensional MnPS$_3$, showing the NN, NNN, and NNNN exchange interactions, denoted by $J_1$, $J_2$, and $J_3$, respectively. The corresponding Brillouin zone and the high-symmetry path are also shown.}
\label{fig:1}
\end{figure}
The three exchange interactions are denoted by $J_1$, $J_2$, and $J_3$, respectively, as shown in Fig.\ref{fig:1}, with their corresponding values tabulated in Table~\ref{table1}.  MnPS$_3$ also exhibits weak uniaxial magnetic anisotropy, with the easy axis oriented perpendicular to the honeycomb layer~\cite{Yan_2024,Wang2023,Kim_2019}. Note that the value of $J$'s are taken from Ref.~\cite{Wildes1998SpinWaves} and there is a factor of 2 difference due to the different conventions used
in the Heisenberg Hamiltonian~\cite{Cheng2016PRL}.
\begin{table}[h!]
\caption{\label{table1}
Exchange interaction, anisotropy energy, and magnetic moments used in the calculations for the two-sublattice ferromagnet (MnPS$_3$).
}
\begin{ruledtabular}
\begin{tabular}{lcc}
Quantity & Value (unit) & Reference \\ 
\colrule
$J_1 \equiv J^{xx}_1 = J^{yy}_1 = J^{zz}_1$ [NN] 
& 1.54\,\, {\rm (meV)} 
& \cite{Wildes1998SpinWaves} \\
$J_2\equiv J^{xx}_2 = J^{yy}_2 = J^{zz}_2$ [NNN] 
& 0.14\,\, {\rm (meV)} 
& \cite{Wildes1998SpinWaves} \\
$J_3\equiv J^{xx}_3 = J^{yy}_3 = J^{zz}_3$ [NNNN] 
& 0.36\,\, {\rm (meV)} 
& \cite{Wildes1998SpinWaves} \\
$K_A = K_B$  
& -0.0086 \,\, {\rm (meV)}
& \cite{Wildes1998SpinWaves} \\
$M^A = M^B$  
& $5.21$\,\,($\mu_B$) & \cite{BABUKA2020109592}\\
\end{tabular}
\end{ruledtabular}
\end{table}

The honeycomb lattice is further characterized by a hexagonal Brillouin zone~\cite{Persoguba2018,Neumann2022}. The center of the Brillouin zone is denoted by $\Gamma$, while $K$ and $K'$ correspond to the inequivalent corners and $M$ to the midpoints of the hexagonal edges. The magnon dispersion is evaluated along the conventional high-symmetry path $\Gamma-K-M-\Gamma$, which captures the essential features of the magnetic excitation spectrum. Such a high-symmetry path is shown in Fig~\ref{fig:1}.

The precessional and inertial motion of magnetic moments can be described by the inertial Landau-Lifshitz-Gilbert (ILLG) equation~\cite{MONDAL_Review}
\begin{align}\label{Eq2}
    \frac{\partial \mathbf{S}_i}{\partial t}  =  \mathbf{S}_i\times\left[-\gamma_i \mathbf{H}^{\rm eff}_i \,+ \,\frac{\alpha_i}{M_{i}}\frac{\partial \mathbf{S}_i}{\partial t} + \frac{\eta_i}{M_i} \frac{\partial^2 \mathbf{S}_i}{\partial t^2}\right]\,,
\end{align}
where $\gamma_i$, $\alpha_i$, and $\eta_i$ are the gyromagnetic ratio, Gilbert damping parameter, and inertial relaxation time, respectively. The effective fields $\mathbf{H}^{\rm eff}_i$ entering in Eq.~\eqref{Eq2} can be computed using the atomistic spin Hamiltonian in Eq.~\eqref{Eq1} via $\mathbf{H}^{\rm eff}_i = -\frac{1}{M_i}\frac{\partial{\mathcal H}}{\partial\bm{S}_i}$.

For a honeycomb antiferromagnet, we consider a N\'eel-ordered ground state in which the magnetic moments on the two sublattices, denoted by $A$ and $B$, are aligned antiparallel along the $+\hat{\mathbf{z}}$ and $-\hat{\mathbf{z}}$ directions, respectively. To describe small-amplitude dynamics around this equilibrium configuration, we introduce the angular variables $\beta_{1i}$ and $\beta_{2i}$ for the spin at site $i\in A$, and $\beta_{1j}$ and $\beta_{2j}$ for the spin at site $j\in B$. The corresponding spin vectors can be expanded as~\cite{Rozsa_2013}
\begin{equation}
\mathbf{S}^{A}_{i}
=
\begin{pmatrix}
\beta_{2i}^A \\
-\beta_{1i}^A \\
1-\dfrac{(\beta_{1i}^A)^{2}+ (\beta_{2i}^A)^{2}}{2}
\end{pmatrix},
\mathbf{S}^{B}_{j}
=
\begin{pmatrix}
\beta_{2j}^B \\
\beta_{1j}^B \\
-1+\dfrac{(\beta_{1j}^B)^{2}+ (\beta_{2j}^B)^{2}}{2}
\end{pmatrix}.
\end{equation}
Substituting these small-deviation expansions into the ILLG equation [Eq.~\eqref{Eq2}] allows us to linearize the equations of motion about the Néel-ordered ground state. We perform a spatial Fourier transformation of the small angular variables, $\widetilde{\beta}_{1,2}({\bf k}_i)$, and introduce the circular components $\widetilde{\beta}_{\pm}(\mathbf{k}_i)=\widetilde{\beta}_{2}(\mathbf{k}_i)\pm i\widetilde{\beta}_{1}(\mathbf{k}_i)$. By assuming harmonic dynamics of the form $\widetilde{\beta}_{\pm}(\mathbf{k}_i,t)\propto e^{i\omega t}$, the time derivatives are replaced by $-i\partial_t\rightarrow\omega$. The linearized ILLG equations can then be expressed as a set of coupled algebraic equations for the Fourier amplitudes. The mathematical procedure follows the framework developed in Ref.~\cite{Mondal2022PRB}. In the present two-sublattice model, the linearized equations of motion
separate into two independent $2\times2$ blocks, corresponding to the
coupled variables
$\left(\widetilde{\beta}_{+}^{A}(\mathbf{k}_i),\widetilde{\beta}_{-}^{B}(\mathbf{k}_i)\right)$
and
$\left(\widetilde{\beta}_{-}^{A}(\mathbf{k}_i),\widetilde{\beta}_{+}^{B}(\mathbf{k}_i)\right)$,
respectively. Denoting the first set by + and the second set by -, these linearised equations can be recast in a matrix form as
\begin{figure*}
    \centering
\includegraphics[width=1\linewidth]{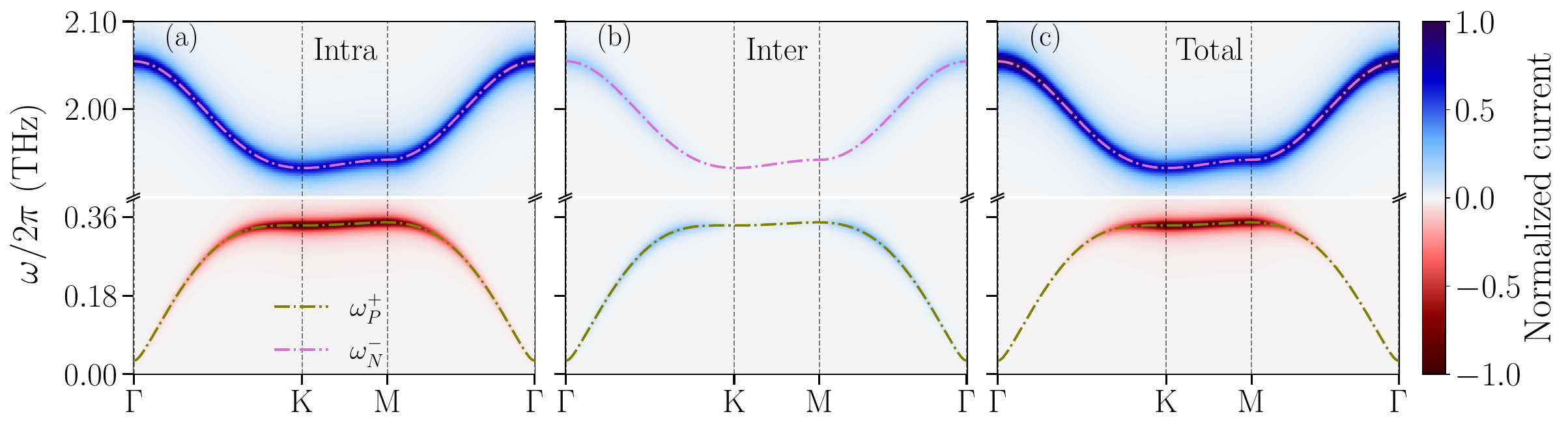}
\caption{Calculated magnon spectra with the weighted (a) intra-sublattice, (b) inter-sublattice, and (c) total spin pumping current along the high-symmetry points in the Brillouin zone.}
    \label{fig2}
\end{figure*}
\begin{widetext}
\begin{align}
\begin{pmatrix}
\widetilde{\beta}_\pm^A({\bf k}_i)
\\
\widetilde{\beta}_\mp^B({\bf k}_i)
\end{pmatrix}
=
\frac{1}{\Delta_{\pm}({\bf k}_i)}
\begin{pmatrix}
-\eta^B\omega^2
\pm i\omega(\alpha^B \pm i)
+\widetilde{\Omega}_B({\bf k}_i)
&
- \widetilde{J}_A(\mathbf{k}_i)
\\
- \widetilde{J}_B(\mathbf{k}_i)
&
-\eta^A\omega^2
\pm i\omega(\alpha^A \mp i)
+\widetilde{\Omega}_A({\bf k}_i)
\end{pmatrix}
\begin{pmatrix}
\mp i\gamma^A\widetilde{B}_{\pm}
\\[10pt]
\mp i\gamma^B\widetilde{B}_{\pm}
\end{pmatrix}
\label{Eq4}
\end{align}
where, 
\begin{align}
\Delta_\pm({\bf k}_i) & = \Bigg[ \Bigg\{ -\eta^A \omega^2 \pm i \omega \left(\alpha^A \mp i \right) + \widetilde{\Omega}_A({\bf k}_i) \Bigg\} \Bigg\{ -\eta^B \omega^2 \pm i \omega \left(\alpha^B \pm i \right) + \widetilde{\Omega}_B({\bf k}_i) \Bigg\} \Bigg] - \Bigg[ \widetilde{J}_A (\mathbf{k}_i) \widetilde{J}_B (\mathbf{k}_i) \Bigg]\\
\widetilde{\Omega}_{A/B}({\bf k}_i) & = \frac{\gamma^{A/B}}{M^{A/B}} \left[ 3 J_1^{zz} - 6 J_2^{zz} + 3 J_3^{zz} + \widetilde{J}_{A/B}^\prime({\bf k}_i) - 2 K_z^{A/B} \pm M^{A/B} B_0 \right]\\
\widetilde{J}_{A/B}({\bf k}_i) & = \frac{\gamma^{A/B}}{M^{A/B}}\left\{J_{1}^{xx} \left[ e^{\pm i \mathbf{k}_i \cdot \boldsymbol{\delta}_1} + e^{\pm i \mathbf{k}_i \cdot \boldsymbol{\delta}_2} + e^{\pm i \mathbf{k}_i \cdot \boldsymbol{\delta}_3} \right] + J_3^{xx} \left[ e^{\pm i \mathbf{k}_i \cdot \mathbf{b}_1} + e^{\pm i \mathbf{k}_i \cdot \mathbf{b}_2} + e^{\pm i \mathbf{k}_i \cdot \mathbf{b}_3} \right]\right\}\\
\widetilde{J}_{A/B}^\prime({\bf k}_i) & = J_2^{xx} \left[ e^{\pm i \mathbf{k}_i \cdot \mathbf{a}_1} + e^{\pm i \mathbf{k}_i \cdot \mathbf{a}_2} + e^{\pm i \mathbf{k}_i \cdot \mathbf{a}_3} + e^{\pm i \mathbf{k}_i \cdot \mathbf{a}_4} + e^{\pm i \mathbf{k}_i \cdot \mathbf{a}_5} + e^{\pm i \mathbf{k}_i \cdot \mathbf{a}_6} \right]
\end{align}
\end{widetext}

The spin pumping current can be calculated via~\cite{Moura2018,Furutani2025,Brataas2022}
\begin{equation}
    j_{\rm sp}=
    \frac{\omega}{2\pi}
    \int_{0}^{2\pi/\omega}
    \frac{\hbar}{4\pi}
    g_r^{\uparrow\downarrow}
    \left[
    \mathbf{S}(t)\times\dot{\mathbf{S}}(t)
    \right]_z
    \,dt,
\end{equation}
where $g_r^{\uparrow\downarrow}$ denotes the real part of the spin-mixing conductance, which we set to $10^{19}$~m$^{-2}$~\cite{Mondal2021PRBSpinCurrent}. 
For the two-sublattice antiferromagnet, the total spin pumping current can be decomposed into intrasublattice and inter-sublattice contributions. The intrasublattice contribution is determined by $
\mathbf{S}^A\times\dot{\mathbf{S}}^A
+
\mathbf{S}^B\times\dot{\mathbf{S}}^B
\equiv
\omega
\left[
\widetilde{\beta}_+^A({\bf k}_i)
\widetilde{\beta}_-^A({\bf k}_i)
+
\widetilde{\beta}_+^B({\bf k}_i)
\widetilde{\beta}_-^B({\bf k}_i)
\right]$,
whereas the inter-sublattice contribution is given by
$\mathbf{S}^A\times\dot{\mathbf{S}}^B
+
\mathbf{S}^B\times\dot{\mathbf{S}}^A
\equiv
\omega
\left[
\widetilde{\beta}_+^A({\bf k}_i)
\widetilde{\beta}_+^B({\bf k}_i)
+
\widetilde{\beta}_-^A({\bf k}_i)
\widetilde{\beta}_-^B({\bf k}_i)
\right]$. The momentum-resolved spin pumping current is obtained by evaluating these two contributions for each magnon mode ${\bf k}_i$ in the Brillouin zone. We therefore evaluate the intrasublattice and inter-sublattice contributions separately along the high-symmetry path of the Brillouin zone to elucidate their respective roles in the spin pumping response. To calculate the spin pumping current, we set $\alpha^A=\alpha^B=\alpha=0.02$ and $\eta^A=\eta^B=\eta=100$~fs. A comparable Gilbert damping parameter has been employed in previous atomistic spin-dynamics simulations of MnPS$_3$~\cite{Alliati2022MnPS3}. The inertial relaxation time is chosen as $\eta=100$~fs, consistent with experimentally measured values of the inertial relaxation time in ferromagnetic materials, which are typically of the order of $100$~fs~\cite{unikandanunni2021inertial}.   

\section{Results and Discussion}
We first investigate the magnon spectrum of the two-dimensional honeycomb antiferromagnet along the high-symmetry path of the magnetic Brillouin zone. Note that the precessional magnon spectrum of MnPS$_3$ has previously been investigated under both weak magnetic fields~\cite{Xing2019,HICKS2019,Olsen_2021,Mai2021,Go2024Nano,Cheng2016PRL,Bazza2021,Wildes2021} and strong magnetic fields~\cite{Nawwar_2025}. Fig.~\ref{fig2} shows the calculated magnon spectrum along the high-symmetry path $\Gamma-K-M-\Gamma$. In the absence of spin inertia, the spectrum consists of the conventional precessional magnon branches, which agree well with previously reported results~\cite{Nawwar_2025,Cheng2016PRL,Bazza2021}. Upon including the inertial term, an additional high-frequency nutational branch emerges, accompanied by a reduction in the energy of the conventional precessional branch. In the absence of an external magnetic field, the symmetry between the two antiparallel magnetic sublattices leads to the degeneracy of the symmetry-related modes. Consequently, the two precessional branches overlap, as do the two nutational branches, resulting in only one distinct precessional and one distinct nutational magnon branch, as evident from Fig.~\ref{fig2}. This degeneracy can be lifted by applying an external magnetic field, as discussed for the precessional modes in Ref.~\cite{Nawwar_2025}. In the present work, we focus on the corresponding inertial dynamics and investigate the nutational modes and, in particular, their contribution to spin pumping. Note that the precessional and nutational eigenmodes correspond to
opposite rotational senses as denoted by $\omega_P^+({\bf k})$ and $\omega_N^-({\bf k})$~\cite{Mondal2020nutation}.

To elucidate the momentum dependence of the spin pumping current, we further calculate $j_{\rm sp}$ 
throughout the Brillouin zone by resolving the contributions from the intra- and inter-sublattice dynamics. The normalized spin pumping current weights are mapped onto the corresponding magnon spectra, as shown in Fig.~\ref{fig2}. We find that the intra-sublattice precessional spin pumping current increases progressively from the $\Gamma$ point toward the Brillouin zone boundary, where it reaches its maximum, while remaining negligibly small near the center of the Brillouin zone. In contrast, the nutational spin pumping current exhibits a much weaker momentum dependence, with its maximum response occurring around the $\Gamma$ point and remaining comparatively robust toward the Brillouin zone boundary. It is worth noting that the precessional and nutational spin pumping
currents exhibit opposite signs, which originates from the opposite
dynamical character of the corresponding eigenmodes~\cite{Mondal2021PRBSpinCurrent}.
\begin{figure}[H]
    \centering
\includegraphics[width=1\linewidth]{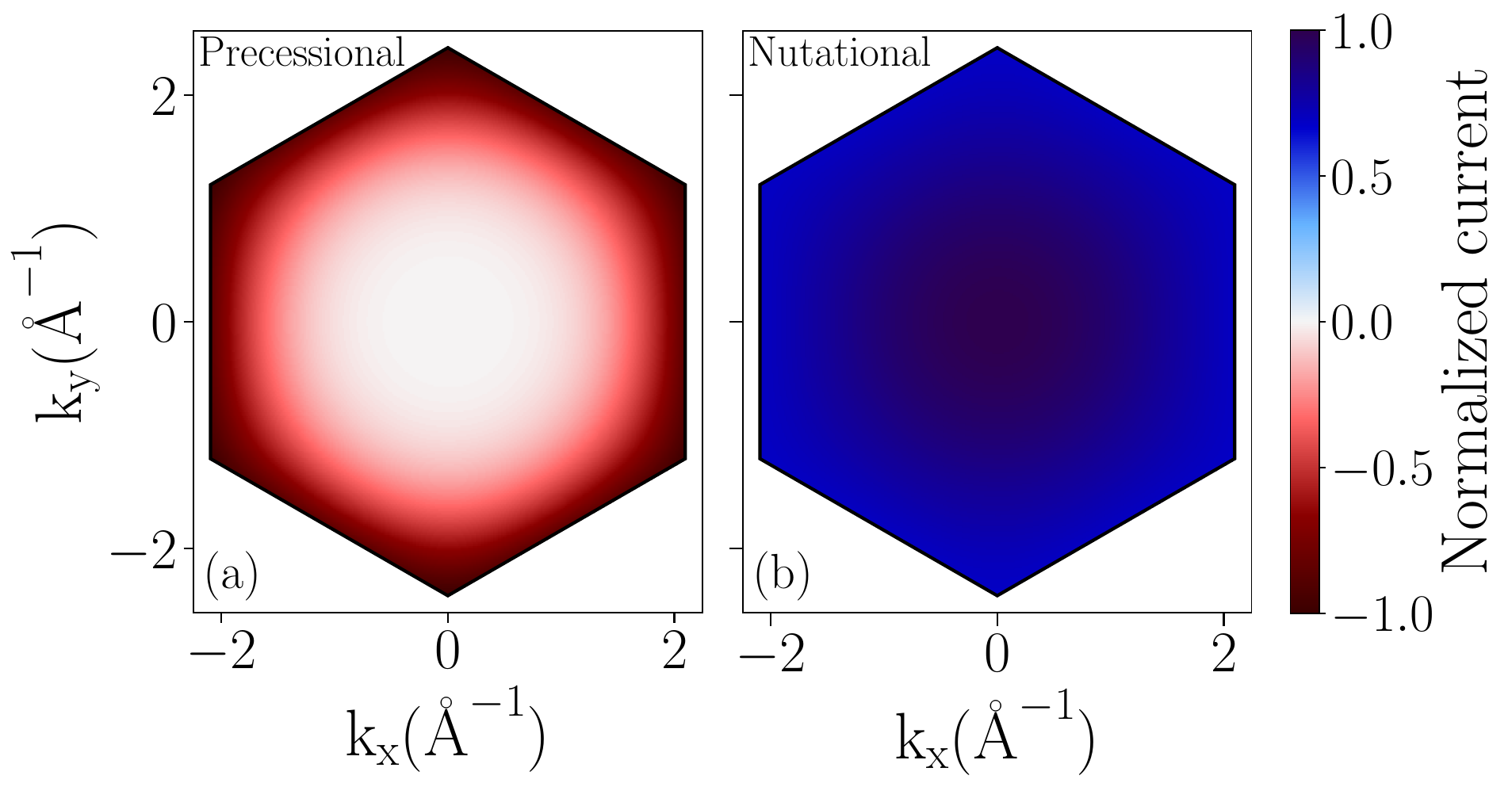}
    \caption{Color maps of the normalized precessional and nutational spin pumping currents as functions of $k_x$ and $k_y$ across the Brillouin zone.}
    \label{fig:3}
\end{figure}
\begin{figure*}
    \centering  
\includegraphics[width=0.9\linewidth]{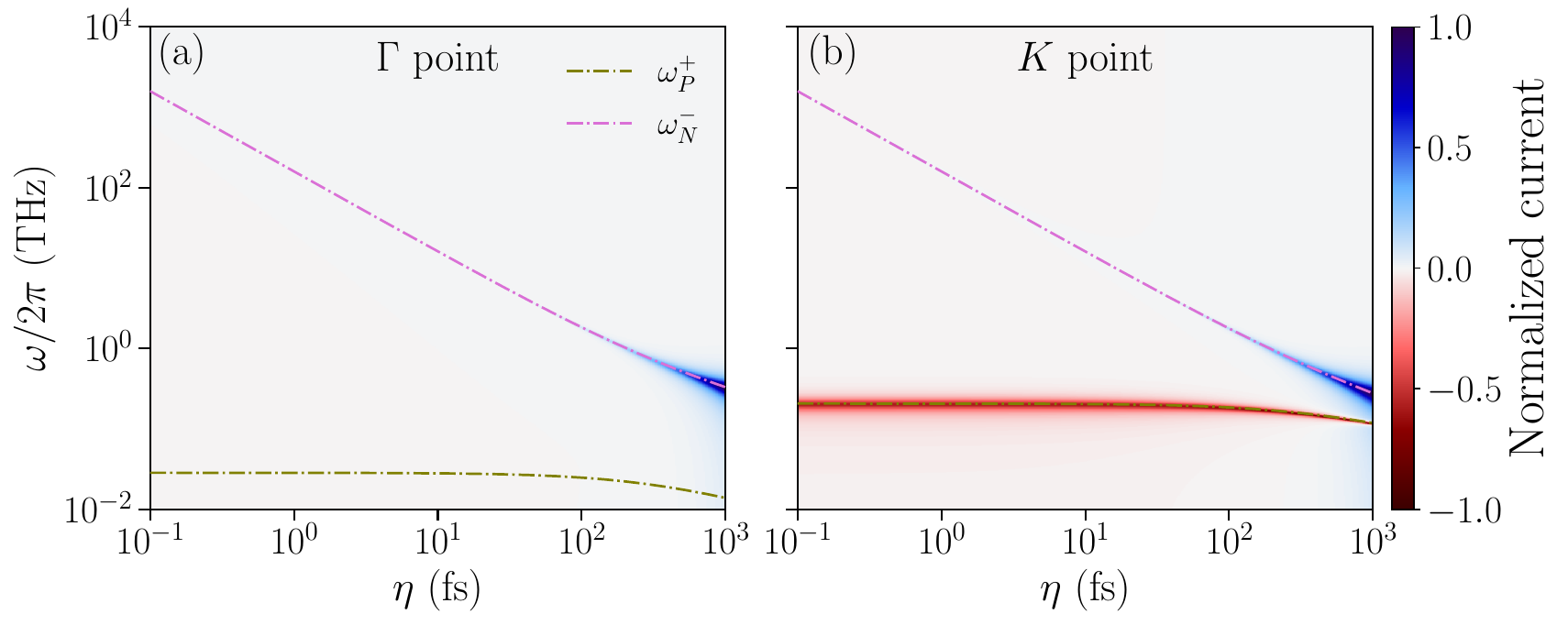}
    \caption{Normalized precessional and nutational spin pumping currents as a
function of the inertial relaxation time $\eta$ at (a) $\Gamma$ and
(b) $K$.}
\label{fig:4}
\end{figure*}

The pronounced enhancement of the intrasublattice precessional spin pumping current toward the Brillouin zone boundary can be understood from the momentum dependence of the intersublattice exchange matrix element. In this case, we find that the leading-order contribution to the spin pumping current is $\sqrt{\widetilde{\Omega}^2({\bf k})-\widetilde{J}^2({\bf k})}
\left(\widetilde{\Omega}^2({\bf k})-\widetilde{J}^2({\bf k})\right)$ (see Appendix~\ref{AppendixA} for detailed calculations). For the honeycomb lattice considered here, we find $\widetilde{J}(\Gamma)\neq 0$, whereas $\widetilde{J}(K)=0$. Thus, the two magnetic sublattices are strongly coupled at the $\Gamma$ point, resulting in a collective precessional eigenmode characterized by nearly equal-amplitude dynamics on the two antiparallel sublattices. Consequently, the spin pumping contributions from the two sublattices largely cancel, leading to a negligibly small precessional spin pumping current at the $\Gamma$ point. Such cancellation is absent at the $K$ point, where the intersublattice coupling vanishes and the two sublattices become effectively decoupled. As a result, the precessional spin pumping current increases toward the Brillouin zone boundary and reaches its maximum near the $K$ point. The intrasublattice exchange matrix elements also contribute to the spin pumping current, since $\widetilde{J}^{\prime}_{A/B}(\Gamma)\neq 0$ and $\widetilde{J}^{\prime}_{A/B}(K)\neq 0$. However, their contribution remains relatively small owing to the weak intrasublattice exchange interaction $J_2$ reported for MnPS$_3$. On the other hand, the nutational spin current relies on the local nature of the inertial dynamics, which are governed primarily by the inertial term. In this case, we find the leading-order contribution to the spin pumping current is $1/\eta^3$. Consequently, the nutational current remains finite even at the $K$ point, where $\widetilde{J}_{A/B}(K)=0$. Additionally, the momentum-dependent nutational spin pumping current arises from higher-order corrections proportional to $\widetilde{J}({\bf k})/\eta^2$ (see Appendix~\ref{AppendixA} for details). As discussed earlier, $\widetilde{J}(\Gamma)\neq0$, whereas $\widetilde{J}(K)=0$. At the $\Gamma$ point, the nutational eigenmodes are dominated by one of the two magnetic sublattices~\cite{Mondal2021PRBSpinCurrent}. Consequently, the spin pumping contributions from the two sublattices do not cancel, in contrast to the precessional modes. The finite $\widetilde{J}(\Gamma)$ therefore enhances the nutational spin pumping current, making it largest around the $\Gamma$ point. Such ${\bf k}$-dependent contributions are absent at the $K$-point because $\widetilde{J}(K)=0$.   

The inter-sublattice spin pumping contribution has been computed in Fig.~\ref{fig2}(b). It exhibits a pronounced momentum dependence as well. The precessional spin current is strongly
suppressed near the $\Gamma$ point and vanishes at the Brillouin zone
boundaries, while remaining finite, although relatively small, at intermediate wave vectors. In this case, we find the leading-order contribution to the spin pumping current is $\widetilde{J}({\bf k})\left(
\widetilde{J}({\bf k})-\widetilde{\Omega}({\bf k})
\right) \sqrt{
\widetilde{\Omega}^{2}({\bf k})
-\widetilde{J}^{2}({\bf k})
}$ (see Appendix~\ref{AppendixA} for details). Hence, at the $K$-point, this current vanishes. More importantly, such
current exhibits a sign opposite to that of the intra-sublattice
contribution throughout the Brillouin zone. This behavior is
consistent with earlier calculations at the $\Gamma$ point
~\cite{Mondal2021PRBSpinCurrent}, while our results demonstrate that
the same sign relation persists over the entire Brillouin zone. On the other hand, the inter-sublattice nutational spin pumping
current remains finite at the $\Gamma$ point, in comparison to its
precessional counterpart. This behavior originates from the dominant
inertial contribution to the nutational response $\widetilde{J}({\bf k})/\eta^2$ (see Appendix~\ref{AppendixA} for details). In contrast, at the Brillouin zone
boundary, the inter-sublattice nutational spin pumping current
vanishes, consistent with our analytical calculations. It is worth noting that the inter-sublattice contribution to the spin pumping
current is generally much smaller than the intra-sublattice contribution
throughout the Brillouin zone, except in the vicinity of the $\Gamma$ point.
Consequently, the total spin pumping current largely follows the momentum
dependence of the intra-sublattice contribution. For the precessional mode, the inter-sublattice contribution has an opposite sign to the
intra-sublattice contribution, leading to a partial cancellation and, therefore, a reduction in the total spin pumping current.
\begin{figure}[H]
   \centering
\includegraphics[width=1\linewidth]{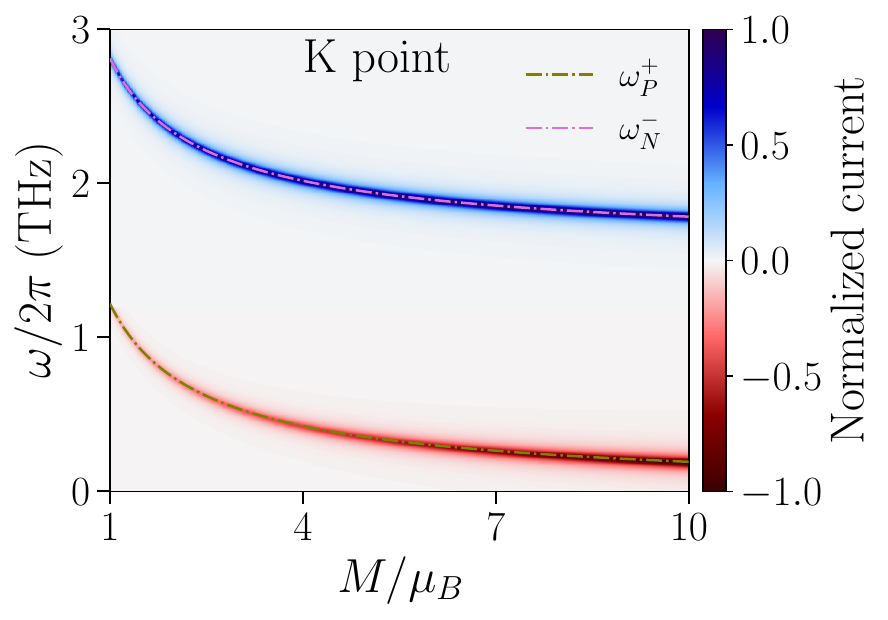}
 \caption{The normalized precessional and nutational spin pumping currents as a function of magnetization at the $K$ point.}
    \label{fig:5}
\end{figure}
A static magnetic field lifts the degeneracy of the magnon bands, resulting in distinct spin pumping currents associated with each nondegenerate magnon band, as discussed in Appendix~\ref{appendixB}.

To further understand the behavior of spin pumping current, we compute the total current, including both intra- and
inter-sublattice contributions, throughout the full Brillouin zone,
$j_{\rm sp}(k_x,k_y)$, as shown in Fig.~\ref{fig:3}. The two-dimensional map reveals a pronounced contrast between the precessional and
nutational responses, as discussed earlier. 
We numerically calculated the amount of spin-pumping current and find that, while the precessional current is nearly absent at the $\Gamma$ point ($0.01\%$) and becomes dominant at the $K$ and $M$ points ($100\%$ and $87\%$, respectively), the nutational current shows the opposite trend, being maximal at $\Gamma$ ($100\%$) and remaining substantial at $K$ and $M$ ($68\%$ and $69\%$, respectively).
This complementary momentum dependence provides a clear distinction between the precessional and nutational spin pumping channels. 

Next, we investigate the dependence of the total spin pumping current on the inertial relaxation time, $\eta$, at the $\Gamma$ and $K$ points, as shown in Fig.~\ref{fig:4}. The overall variation of the magnon frequency with $\eta$ is in good agreement with earlier reports~\cite{Mondal2020nutation}. As discussed above, the precessional spin pumping current is strongly suppressed at the $\Gamma$ point, and this suppression remains essentially unchanged with increasing $\eta$. In contrast, at the $K$ point, the precessional spin pumping current exhibits a pronounced dependence on $\eta$, decreasing as $\eta$ increases. This is reasonable to expect as the correction term due to inertia scales with $-\eta\widetilde{J}({\bf k})\left(\widetilde{J}({\bf k})+\widetilde{\Omega}({\bf k})\right)$. The nutational spin pumping current shows the opposite trend, increasing with $\eta$ at both the $\Gamma$ and $K$ points because such a current is proportional to $1/\eta^3$. Notably, this enhancement is substantially stronger at the $K$ point than at the $\Gamma$ point.

The spin pumping current also exhibits a dependence on the magnetization. At the $\Gamma$ point, however, this dependence is relatively weak because the precessional current is strongly suppressed, while the nutational current remains robust. In contrast, at the $K$ point, although the nutational current remains appreciable, the precessional current increases with increasing magnetization. This magnetization dependence of the spin pumping current at the $K$ point is shown in Fig.~\ref{fig:5}. Experimentally, the $K$ point can be advantageous for detecting nutational spin pumping at lower magnetization, where the precessional contribution is suppressed. Thus, controlling the magnetization provides a practical route to enhancing the relative contribution of the nutational spin pumping current.    

\begin{figure}
   \centering
\includegraphics[width=1\linewidth]{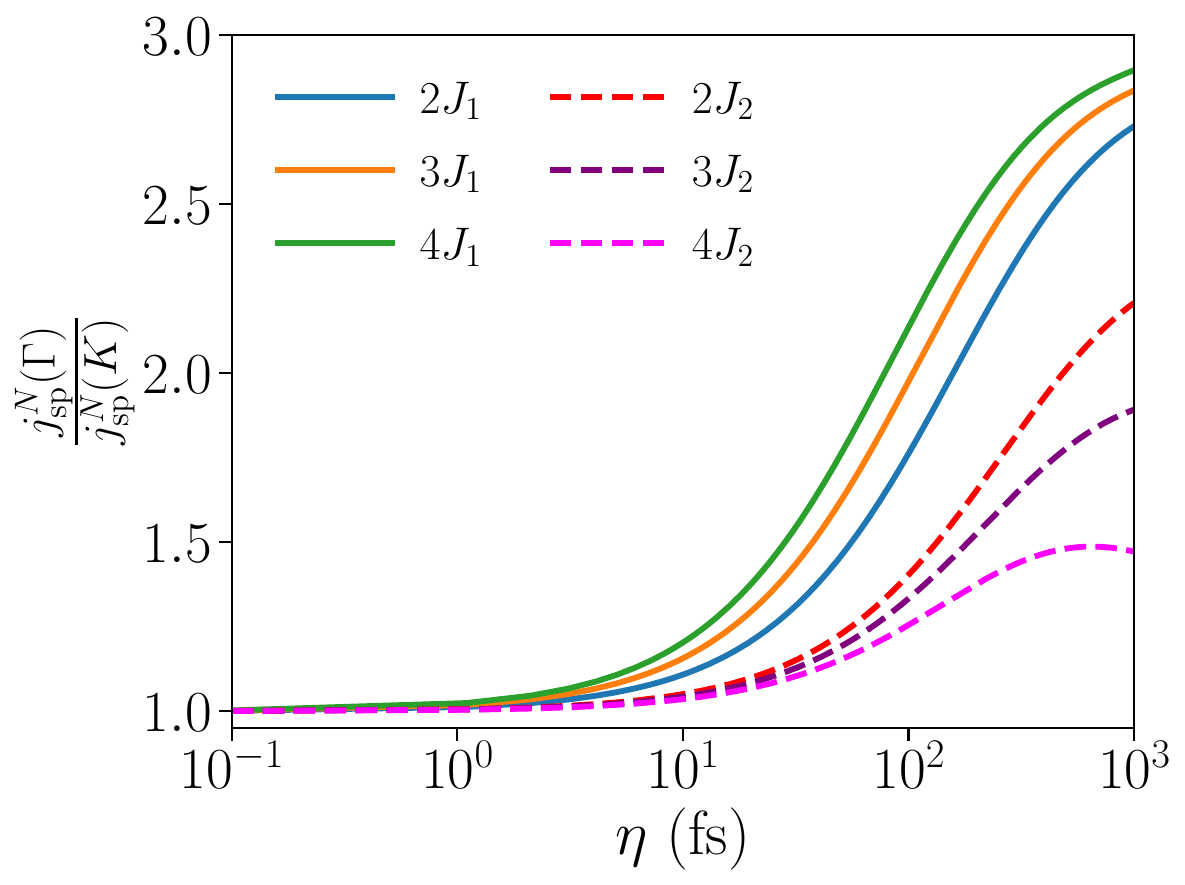}
 \caption{The ratio of nutational current at the $\Gamma$-point to the $K$-point as a function of $\eta$ at several NN and NNN exchange strengths.}
    \label{fig:6}
\end{figure}

Next, we focus on the nutational spin pumping current and calculate the ratio of currents at the $\Gamma$ point to that at the $K$ point as a function of the inertial relaxation time $\eta$ for several values of the nearest-neighbor ($J_1$), and next-nearest neighbor ($J_2$) exchange interaction, as shown in Fig.~\ref{fig:6}. Note that the effect of $J_3$ will be similar to $J_1$. 
For $\eta < 10$ fs, this ratio approaches unity for all considered
values of $J_1$ and $J_2$. We find that the nutational current increases with the inertial relaxation time $\eta$ at both the $\Gamma$ and $K$ points. However, the rate of increase is significantly higher at the $\Gamma$ point than at the $K$ point. Therefore, the nutational current ratio increases with $\eta$. This trend is most pronounced in the intermediate range of $\eta \sim 10$ to $100$ fs. In contrast, for smaller relaxation times ($\eta < 10$ fs) and for values approaching $\eta \sim 1$ ps, the rate of increase becomes relatively weak, resulting in a much slower variation of the current ratio.

Finally, we propose a possible experimental scenario and material platform to validate such results. The momentum-sensitive probes, such as Brillouin light scattering~\cite{Demo2001,Sebastian2015} or time-resolved magneto-optical measurements~\cite{Lyalin2023,Stamm2017}, can be used to map the momentum-dependent magnon dynamics with the measured spin pumping signal. In particular, the predicted enhancement of the nutational spin pumping current near the $\Gamma$ point provides a direct experimental signature that distinguishes it from the conventional precessional response. Furthermore, magnetic systems with a larger inertial relaxation time $\eta$ are expected to exhibit enhanced nutational spin pumping currents accompanied by a suppression of the precessional contribution. Although the value of $\eta$ remains largely unexplored in many magnetic materials, including two-dimensional van der Waals magnets, recent studies have suggested that the inertial relaxation time can be modified through spin-orbit coupling~\cite{unikandanunni2021inertial}. This provides a possible route for engineering the inertial response by interfacing a two-dimensional antiferromagnet with a heavy-metal layer possessing strong spin-orbit coupling. Such a heterostructure has already been employed to investigate for the detection of magnon-mediated spin signals~\cite{Shiomi2017,Feringa2022,Xing2019}. For a two-dimensional antiferromagnet, an increase in the antiferromagnetic exchange coupling enhances the nutational spin pumping current while simultaneously suppressing the precessional contribution. Thus, an antiferromagnet with stronger nearest-neighbor antiferromagnetic coupling is favored.

\section{conclusion}
In summary, we have investigated momentum-resolved spin pumping in a two-dimensional honeycomb antiferromagnet by incorporating spin inertia into the Landau-Lifshitz-Gilbert equation. Using a microscopic two-sublattice model with $J_1$-$J_2$-$J_3$ exchange interactions, we have obtained the precessional and nutational magnon modes and evaluated their corresponding intra- and inter-sublattice spin pumping contributions throughout the Brillouin zone. Our results reveal a pronounced contrast between the momentum dependence of the two types of dynamics. While the precessional spin pumping current is strongly suppressed near the $\Gamma$ point and increases towards the Brillouin zone boundary, the nutational current reaches its maximum around $\Gamma$ and decreases towards the zone boundary. 

We further demonstrate that the nutational spin pumping response is strongly governed by the inertial relaxation time $\eta$. In the small-$\eta$ regime, the nutational current is dominated by a leading contribution proportional to $1/\eta^3$, whereas the momentum-dependent corrections enter at higher orders in $\eta$. Consequently, the momentum dependence becomes increasingly pronounced as $\eta$ increases, leading to a stronger enhancement of the nutational current near the $\Gamma$ point relative to the $K$ point. In contrast, the precessional spin pumping current exhibits a comparatively weak dependence on $\eta$. We have also computed the dependence of the spin pumping response on the magnetic moment and exchange interactions. The application of a static magnetic field lifts the degeneracy of the magnon bands. We find that the field redistributes the spin pumping current among the nondegenerate magnon branches, with the ratio of the nutational currents of the two branches increasing with magnetic field, while the corresponding precessional ratio decreases.

Finally, our results suggest a feasible route towards experimental observation of contrasting momentum-dependent precessional and nutational spin pumping. An antiferromagnet characterized by strong exchange coupling, a small magnetic moment, and a large inertial relaxation time $\eta$ is particularly promising for experimentally observing the enhanced nutational and suppressed precessional spin pumping responses predicted here. 


\section{ACKNOWLEDGMENTS}
We acknowledge funding support from the SERB-SRG via Project No. SRG/2023/000612 and the faculty research scheme at IIT (ISM) Dhanbad, India, under Project No. FRS(196)/2023-2024/PHYSICS.

\appendix
\setcounter{figure}{0}
\renewcommand{\thefigure}{A.\arabic{figure}}
\section{Calculations of intra- and inter-sublattice spin pumping currents}
\label{AppendixA}
\begin{figure*}
   \centering
\includegraphics[width=1\linewidth]{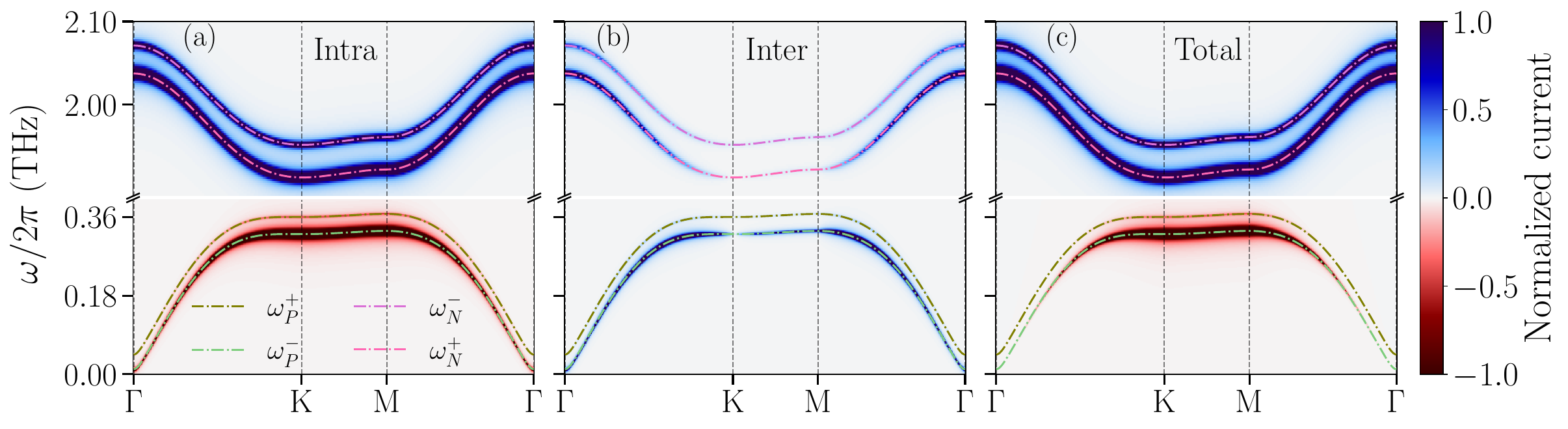}
 \caption{Calculated magnon spectra with the weighted (a) intra-sublattice, (b) inter-sublattice, and (c) total spin current along the high-symmetry points in the Brillouin zone. The magnetic field used in the calculation is $B_0 = 1$ T.}
    \label{fig:S1}
\end{figure*}
To analytically calculate the spin pumping current, we assume that $\alpha^A = \alpha^B = \alpha = 0$ and $B_0 = 0$. Further, assuming $\eta^A = \eta^B = \eta$, $\gamma^A = \gamma^B = \gamma$, and $M^A = M^B = M$, we find $\widetilde{\Omega}_A({\bf k}) =\widetilde{\Omega}_B({\bf k}) = \widetilde{\Omega}({\bf k})$, $\widetilde{J}_A({\bf k}) =\widetilde{J}_B({\bf k}) = \widetilde{J}({\bf k})$ and $\Delta_+({\bf k}) = \Delta_-({\bf k}) = \Delta({\bf k})$. Thus, we can write four linearised equations following Eq.~\eqref{Eq4}
as
\begin{align}
\widetilde{\beta}_+^A({\bf k})
&=
-\frac{i\gamma\widetilde{B}_+}{\Delta({\bf k})}
\left[
\widetilde{\Omega}({\bf k})
-\eta\omega^2
-\omega
-\widetilde{J}({\bf k})
\right],
\\
\widetilde{\beta}_-^B({\bf k})
&=
-\frac{i\gamma\widetilde{B}_+}{\Delta({\bf k})}
\left[
\widetilde{\Omega}({\bf k})
-\eta\omega^2
+\omega
-\widetilde{J}({\bf k})
\right],
\\
\widetilde{\beta}_-^A({\bf k})
&=
\frac{i\gamma\widetilde{B}_-}{\Delta({\bf k})}
\left[
\widetilde{\Omega}({\bf k})
-\eta\omega^2
-\omega
-\widetilde{J}({\bf k})
\right],
\\
\widetilde{\beta}_+^B({\bf k})
&=
\frac{i\gamma\widetilde{B}_-}{\Delta({\bf k})}
\left[
\widetilde{\Omega}({\bf k})
-\eta\omega^2
+\omega
-\widetilde{J}({\bf k})
\right].
\end{align}
The $\bf k$-dependent precessional and nutation frequencies can be calculated by setting $\Delta({\bf k}) = 0$. These frequencies are
\begin{align}
    \omega_P^2 & =
\frac{
1+2\eta\widetilde{\Omega}({\bf k})-D({\bf k})
}
{2\eta^2}
\\
\omega_N^2
& =
\frac{
1+2\eta\widetilde{\Omega}({\bf k})+D({\bf k})
}
{2\eta^2}
\end{align}
where $D({\bf k}) = \sqrt{1+4\eta\widetilde{\Omega}({\bf k}) + 4\eta^2\widetilde{J}^2({\bf k})}$. 
Consequently, the intra- and inter-sublattice spin pumping current can be computed as follows
\begin{align}
    j_{\rm sp}^{\rm intra} & \equiv \omega\left[\widetilde{\beta}_+^A({\bf k})
\widetilde{\beta}_-^A({\bf k})
+
\widetilde{\beta}_+^B({\bf k})
\widetilde{\beta}_-^B({\bf k})\right]\nonumber\\
& =\frac{2\gamma^2\widetilde{B}_+\widetilde{B}_-}{\Delta^2({\bf k})}\omega
\,\mathcal{C}_{\mathrm{intra}}\\
j_{\rm sp}^{\rm inter} & \equiv \omega\left[  \widetilde{\beta}_+^A({\bf k})
\widetilde{\beta}_+^B({\bf k})
+
\widetilde{\beta}_-^A({\bf k})
\widetilde{\beta}_-^B({\bf k})\right] \nonumber\\
& = \frac{2\gamma^2\widetilde{B}_+\widetilde{B}_-}{\Delta^2({\bf k})}\omega 
\,\mathcal{C}_{\mathrm{inter}}
\end{align}
where $\mathcal{C}_{\mathrm{intra}} = \left\{\widetilde{\Omega}({\bf k})
-\eta\omega^2-\widetilde{J}({\bf k})
\right\}^2+\omega^2$ and $\mathcal{C}_{\mathrm{intra}}=\mathcal{C}_{\mathrm{inter}}+2\omega^2$.
For precessional and nutational magnons, neglecting the higher-order correction terms, the intra- and inter-sublattice contributions are calculated as
\begin{widetext}
\begin{align}
\omega_P \,\mathcal{C}_{\mathrm{intra}}^{P}
&\approx
2\left(
\widetilde{J}({\bf k})-\widetilde{\Omega}({\bf k})
\right)\sqrt{\widetilde{\Omega}^2({\bf k})-\widetilde{J}^2({\bf k})}
\Bigg[
-\left(\widetilde{J}({\bf k}) +\widetilde{\Omega}({\bf k})\right)
+\widetilde{J}({\bf k})
\left\{
1-\eta
\left(\widetilde{J}({\bf k})+\widetilde{\Omega}({\bf k})\right)
\left(1-2\eta\widetilde{\Omega}({\bf k})\right)
\right\}
\Bigg]\\
\omega_P \,\mathcal{C}_{\mathrm{inter}}^{P}
&\approx
 2 \left(
\widetilde{J}({\bf k})-\widetilde{\Omega}({\bf k})
\right) \sqrt{
\widetilde{\Omega}^{2}({\bf k})
-\widetilde{J}^{2}({\bf k})
}\,\,
\widetilde{J}({\bf k})\Bigg[
1-\eta \left(
\widetilde{J}({\bf k})+\widetilde{\Omega}({\bf k}) \right)
\left( 1 - 2\eta \widetilde{\Omega}({\bf k}) \right)
\Bigg]\end{align}
\begin{align}
{\omega_N}
\,\mathcal{C}_{\mathrm{intra}}^N
&\approx
{2}
\left[
\frac{1}{\eta^3}
+
\widetilde{J}({\bf k})
\left( \frac{1}{\eta^2} + \frac{1}{\eta} \left( \widetilde{J}({\bf k}) + \widetilde{\Omega}({\bf k}) \right) \right)
\right]\\
{\omega_N}
\,\mathcal{C}_{\mathrm{inter}}^N
&\approx {2\widetilde{J}({\bf k})}
\left[
\frac{1}{\eta^2}
+
\frac{1}{\eta}
\left(
\widetilde{J}({\bf k})+\widetilde{\Omega}({\bf k})
\right)
\right]
\end{align}
\end{widetext}

\begin{figure}
   \centering
\includegraphics[width=1\linewidth]{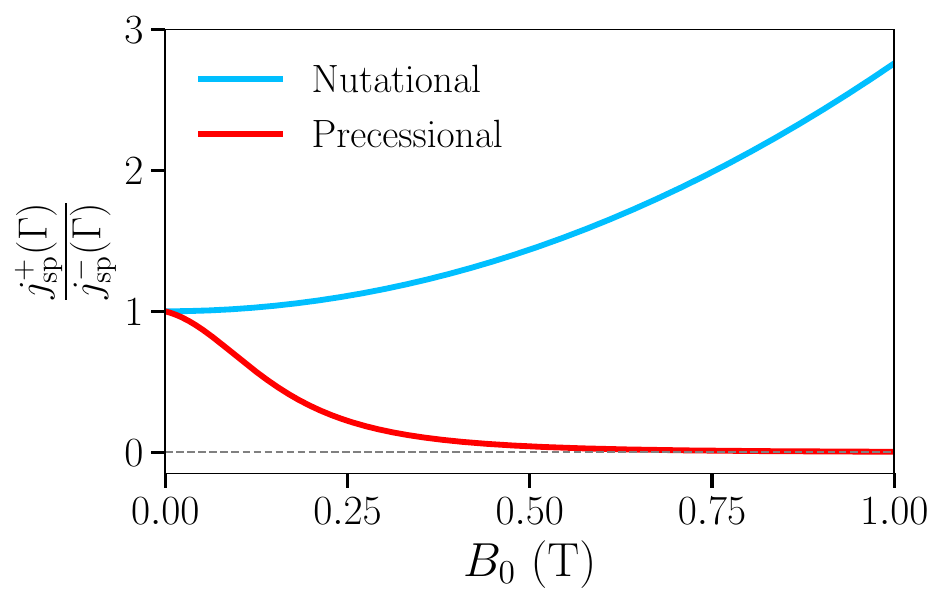}
 \caption{Computed spin pumping current ratio of the nondegenerate magnon bands at the $\Gamma$-point as a function of magnetic field.}
\label{fig:S2}
\end{figure}
Note that while expanding $D({\bf k})$, the higher-order terms have been dropped. These approximate expressions become useful in explaining the results presented in the main text.  

\section{Effect of magnetic field on the spin current}
\label{appendixB}
The application of an external magnetic field breaks the symmetry
between the two antiparallel magnetic sublattices, thereby lifting the
degeneracy of the magnon bands. This lifting of the degeneracy is also
reflected in our spin pumping current calculations for the honeycomb
antiferromagnet MnPS$_3$, as shown in Fig.~\ref{fig:S1}. Our obtained magnon bands
are consistent with previous calculations of the field-dependent
magnon spectrum~\cite{Nawwar_2025}. Note that the non-degenerate magnon bands will have opposite polarity e.g., $\omega_P^+$ and $\omega_P^-$ for the precessional magnon bands, and the corresponding nutational bands will be denoted as $\omega_N^-$ and $\omega_N^+$. Due to the nonzero magnetic field, the spin current contribution changes across the Brillouin zone. Importantly, the spin current contributions do not split equally in the two non-degenerate magnon bands. However, the inter-sublattice contribution remains smaller compared to the intra-sublattice counterpart. Due to opposite precessional spin current for intra- and inter-sublattice contributions, the total current largely remains dominant around the edges of the Brillouin zone.  

The non-equivalent weight of the spin current can further be understood via the computation of the ratio of spin currents at the non-degenerate magnon bands i.e., $j_{\rm sp}^+/j_{\rm sp}^-$, as a function of external magnetic field strength $B_0$ in Fig.~\ref{fig:S2}. At the $\Gamma$-point, this ratio for the precessional spin pumping current decreases with increasing magnetic field and becomes very small, indicating that the high-frequency nondegenerate magnon band contributes negligibly to the spin pumping current. In contrast, the corresponding ratio for the nutational spin pumping current increases with increasing magnetic field.

\bibliography{References}

@article{Stamm2017,
  title = {{Magneto-Optical Detection of the Spin Hall Effect in Pt and W Thin Films}},
  author = {Stamm, C. and Murer, C. and Berritta, M. and Feng, J. and Gabureac, M. and Oppeneer, P. M. and Gambardella, P.},
  journal = {Phys. Rev. Lett.},
  volume = {119},
  issue = {8},
  pages = {087203},
  numpages = {6},
  year = {2017},
  month = {Aug},
  publisher = {American Physical Society},
  doi = {10.1103/PhysRevLett.119.087203},
  url = {https://link.aps.org/doi/10.1103/PhysRevLett.119.087203}
}

@ARTICLE{Sebastian2015,
AUTHOR={Sebastian, Thomas  and Schultheiss, Katrin  and Obry, Björn  and Hillebrands, Burkard  and Schultheiss, Helmut },
TITLE={Micro-focused Brillouin light scattering: imaging spin waves at the nanoscale},
          
JOURNAL={Frontiers in Physics},
          
VOLUME={3},
  
YEAR={2015},
  
URL={https://www.frontiersin.org/journals/physics/articles/10.3389/fphy.2015.00035},
  
DOI={10.3389/fphy.2015.00035}}

@article{Demo2001,
title = {Brillouin light scattering studies of confined spin waves: linear and nonlinear confinement},
journal = {Physics Reports},
volume = {348},
number = {6},
pages = {441-489},
year = {2001},
issn = {0370-1573},
doi = {https://doi.org/10.1016/S0370-1573(00)00116-2},
url = {https://www.sciencedirect.com/science/article/pii/S0370157300001162},
author = {S.O. Demokritov and B. Hillebrands and A.N. Slavin}}

@article{Shiomi2017,
  title = {{Experimental evidence consistent with a magnon Nernst effect in the antiferromagnetic insulator ${\mathrm{MnPS}}_{3}$}},
  author = {Shiomi, Y. and Takashima, R. and Saitoh, E.},
  journal = {Phys. Rev. B},
  volume = {96},
  issue = {13},
  pages = {134425},
  numpages = {9},
  year = {2017},
  month = {Oct},
  publisher = {American Physical Society},
  doi = {10.1103/PhysRevB.96.134425},
  url = {https://link.aps.org/doi/10.1103/PhysRevB.96.134425}
}

@article{Feringa2022,
  title = {{Spin-flop transition in the quasi-two-dimensional antiferromagnet ${\mathrm{MnPS}}_{3}$ detected via thermally generated magnon transport}},
  author = {Feringa, F. and Vink, J. M. and van Wees, B. J.},
  journal = {Phys. Rev. B},
  volume = {106},
  issue = {22},
  pages = {224409},
  numpages = {10},
  year = {2022},
  month = {Dec},
  publisher = {American Physical Society},
  doi = {10.1103/PhysRevB.106.224409},
  url = {https://link.aps.org/doi/10.1103/PhysRevB.106.224409}
}

@article{Lyalin2023,
  title = {{Magneto-Optical Detection of the Orbital Hall Effect in Chromium}},
  author = {Lyalin, Igor and Alikhah, Sanaz and Berritta, Marco and Oppeneer, Peter M. and Kawakami, Roland K.},
  journal = {Phys. Rev. Lett.},
  volume = {131},
  issue = {15},
  pages = {156702},
  numpages = {6},
  year = {2023},
  month = {Oct},
  publisher = {American Physical Society},
  doi = {10.1103/PhysRevLett.131.156702},
  url = {https://link.aps.org/doi/10.1103/PhysRevLett.131.156702}
}

@article{Brataas2022,
  title = {Quantum scattering theory of spin transfer torque, spin pumping, and fluctuations},
  author = {Brataas, Arne},
  journal = {Phys. Rev. B},
  volume = {106},
  issue = {6},
  pages = {064402},
  numpages = {19},
  year = {2022},
  month = {Aug},
  publisher = {American Physical Society},
  doi = {10.1103/PhysRevB.106.064402},
  url = {https://link.aps.org/doi/10.1103/PhysRevB.106.064402}
}

@article{Furutani2025,
  title = {Ginzburg-Landau theory of spin pumping through an antiferromagnetic layer near the N\'eel temperature},
  author = {Furutani, Yuto and Fukushima, Hayato and Yamamoto, Yutaka and Ichioka, Masanori and Adachi, Hiroto},
  journal = {Phys. Rev. B},
  volume = {112},
  issue = {17},
  pages = {174416},
  numpages = {12},
  year = {2025},
  month = {Nov},
  publisher = {American Physical Society},
  doi = {10.1103/yntc-7fvx},
  url = {https://link.aps.org/doi/10.1103/yntc-7fvx}
}

@article{Moura2018,
  title = {Theoretical analysis of FMR-driven spin pumping current and its properties via the self-consistent harmonic approximation},
  author = {Moura, A. R.},
  journal = {Phys. Rev. B},
  volume = {106},
  issue = {5},
  pages = {054313},
  numpages = {11},
  year = {2022},
  month = {Aug},
  publisher = {American Physical Society},
  doi = {10.1103/PhysRevB.106.054313},
  url = {https://link.aps.org/doi/10.1103/PhysRevB.106.054313}
}

@article{Neumann2022,
  title = {Thermal Hall Effect of Magnons in Collinear Antiferromagnetic Insulators: Signatures of Magnetic and Topological Phase Transitions},
  author = {Neumann, Robin R. and Mook, Alexander and Henk, J\"urgen and Mertig, Ingrid},
  journal = {Phys. Rev. Lett.},
  volume = {128},
  issue = {11},
  pages = {117201},
  numpages = {8},
  year = {2022},
  month = {Mar},
  publisher = {American Physical Society},
  doi = {10.1103/PhysRevLett.128.117201},
  url = {https://link.aps.org/doi/10.1103/PhysRevLett.128.117201}
}

@article{Persoguba2018,
  title = {Dirac Magnons in Honeycomb Ferromagnets},
  author = {Pershoguba, Sergey S. and Banerjee, Saikat and Lashley, J. C. and Park, Jihwey and \AA{}gren, Hans and Aeppli, Gabriel and Balatsky, Alexander V.},
  journal = {Phys. Rev. X},
  volume = {8},
  issue = {1},
  pages = {011010},
  numpages = {8},
  year = {2018},
  month = {Jan},
  publisher = {American Physical Society},
  doi = {10.1103/PhysRevX.8.011010},
  url = {https://link.aps.org/doi/10.1103/PhysRevX.8.011010}
}

@article{GAN2026174373,
title = {The effects of magnons in spin pumping},
journal = {Journal of Magnetism and Magnetic Materials},
volume = {655},
pages = {174373},
year = {2026},
issn = {0304-8853},
doi = {https://doi.org/10.1016/j.jmmm.2026.174373},
url = {https://www.sciencedirect.com/science/article/pii/S0304885326005652},
author = {Ling Gan and Shufeng Zhang}
}

@article{Go2024Nano,
    author = {Go, Gyungchoon and An, Daehyeon and Lee, Hyun-Woo and Kim, Se Kwon},
    title = {Magnon Orbital
Nernst Effect in Honeycomb Antiferromagnets
without Spin–Orbit Coupling},
    journal = {Nano Letters},
    volume = {24},
    number = {20},
    pages = {5968-5974},
    year = {2024},
    month = {04},
    issn = {1530-6984},
    doi = {10.1021/acs.nanolett.4c00430},
    url = {https://doi.org/10.1021/acs.nanolett.4c00430},
}

@article{Olsen_2021,
doi = {10.1088/1361-6463/ac000e},
url = {https://doi.org/10.1088/1361-6463/ac000e},
year = {2021},
month = {may},
publisher = {IOP Publishing},
volume = {54},
number = {31},
pages = {314001},
author = {Olsen, Thomas},
title = {{Magnetic anisotropy and exchange interactions of two-dimensional FePS$_3$, NiPS$_3$ and MnPS$_3$ from first principles calculations}},
journal = {Journal of Physics D: Applied Physics}
}

@article{Mai2021,
author = {Thuc T. Mai  and Kevin F. Garrity  and Amber McCreary  and Joshua Argo  and Jeffrey R. Simpson  and Vicky Doan-Nguyen  and Rolando Valdés Aguilar  and Angela R. Hight Walker },
title = {{Magnon-phonon hybridization in 2D antiferromagnet MnPSe$_3$}},
journal = {Science Advances},
volume = {7},
number = {44},
pages = {eabj3106},
year = {2021},
doi = {10.1126/sciadv.abj3106},
URL = {https://www.science.org/doi/abs/10.1126/sciadv.abj3106}}

@article{Wildes2021,
  title = {{Search for nonreciprocal magnons in ${\mathrm{MnPS}}_{3}$}},
  author = {Wildes, A. R. and Okamoto, S. and Xiao, D.},
  journal = {Phys. Rev. B},
  volume = {103},
  issue = {2},
  pages = {024424},
  numpages = {6},
  year = {2021},
  month = {Jan},
  publisher = {American Physical Society},
  doi = {10.1103/PhysRevB.103.024424},
  url = {https://link.aps.org/doi/10.1103/PhysRevB.103.024424}
}

@article{Bazza2021,
  title = {Magnetoelastic coupling enabled tunability of magnon spin current generation in two-dimensional antiferromagnets},
  author = {Bazazzadeh, N. and Hamdi, M. and Park, S. and Khavasi, A. and Mohseni, S. M. and Sadeghi, A.},
  journal = {Phys. Rev. B},
  volume = {104},
  issue = {18},
  pages = {L180402},
  numpages = {6},
  year = {2021},
  month = {Nov},
  publisher = {American Physical Society},
  doi = {10.1103/PhysRevB.104.L180402},
  url = {https://link.aps.org/doi/10.1103/PhysRevB.104.L180402}
}

@article{Cheng2016PRL,
  title = {Spin Nernst Effect of Magnons in Collinear Antiferromagnets},
  author = {Cheng, Ran and Okamoto, Satoshi and Xiao, Di},
  journal = {Phys. Rev. Lett.},
  volume = {117},
  issue = {21},
  pages = {217202},
  numpages = {5},
  year = {2016},
  month = {Nov},
  publisher = {American Physical Society},
  doi = {10.1103/PhysRevLett.117.217202},
  url = {https://link.aps.org/doi/10.1103/PhysRevLett.117.217202}
}

@article{HICKS2019,
title = {Magnetic dipole splitting of magnon bands in a two dimensional antiferromagnet},
journal = {Journal of Magnetism and Magnetic Materials},
volume = {474},
pages = {512-516},
year = {2019},
issn = {0304-8853},
doi = {https://doi.org/10.1016/j.jmmm.2018.10.136},
url = {https://www.sciencedirect.com/science/article/pii/S0304885318306723},
author = {T.J. Hicks and T. Keller and A.R. Wildes}
}

@article{Nawwar_2025,
doi = {10.1088/1361-6633/adf916},
url = {https://doi.org/10.1088/1361-6633/adf916},
year = {2025},
month = {aug},
publisher = {IOP Publishing},
volume = {88},
number = {8},
pages = {080503},
author = {Nawwar, Mohamed and Neumann, Robin R and Wen, Jiamin and Mertig, Ingrid and Mook, Alexander and Heremans, Joseph P},
title = {{Large thermal Hall effect in MnPS$_3$}},
journal = {Reports on Progress in Physics}
}

@article{Xing2019,
  title = {Magnon Transport in Quasi-Two-Dimensional van der Waals Antiferromagnets},
  author = {Xing, Wenyu and Qiu, Luyi and Wang, Xirui and Yao, Yunyan and Ma, Yang and Cai, Ranran and Jia, Shuang and Xie, X. C. and Han, Wei},
  journal = {Phys. Rev. X},
  volume = {9},
  issue = {1},
  pages = {011026},
  numpages = {7},
  year = {2019},
  month = {Feb},
  publisher = {American Physical Society},
  doi = {10.1103/PhysRevX.9.011026},
  url = {https://link.aps.org/doi/10.1103/PhysRevX.9.011026}
}

@article{Liu2021APL,
  author    = {Qi Liu and Y. Zhang and L. Sun and Bingfeng Miao and X. R. Wang and H. F. Ding},
  title     = {Influence of the Spin Pumping Induced Inverse Spin Hall Effect on Spin-Torque Ferromagnetic Resonance Measurements},
  journal   = {Applied Physics Letters},
  volume    = {118},
  number    = {13},
  pages     = {132401},
  year      = {2021},
  doi       = {10.1063/5.0043035},
  url       = {https://doi.org/10.1063/5.0043035},
  publisher = {AIP Publishing}
}

@article{Quarenta2024,
  title = {Bath-Induced Spin Inertia},
  author = {Quarenta, Mario Gaspar and Tharmalingam, Mithuss and Ludwig, Tim and Yuan, H. Y. and Karwacki, Lukasz and Verstraten, Robin C. and Duine, Rembert A.},
  journal = {Phys. Rev. Lett.},
  volume = {133},
  issue = {13},
  pages = {136701},
  numpages = {7},
  year = {2024},
  month = {Sep},
  publisher = {American Physical Society},
  doi = {10.1103/PhysRevLett.133.136701},
  url = {https://link.aps.org/doi/10.1103/PhysRevLett.133.136701}
}

@article{Moussa2026,
  title = {Unquenched orbital angular momentum as the origin of spin inertia},
  author = {Moussa, Tarek and Basu, Darpa Narayan and Mondal, Ritwik and Kamra, Akashdeep},
  journal = {Phys. Rev. B},
  volume = {114},
  issue = {2},
  pages = {024424},
  numpages = {16},
  year = {2026},
  month = {Jul},
  publisher = {American Physical Society},
  doi = {10.1103/9h9p-8qzd},
  url = {https://link.aps.org/doi/10.1103/9h9p-8qzd}
}

@article{Rimmler2025,
  author    = {Berthold H. Rimmler and Banabir Pal and Stuart S. P. Parkin},
  title     = {Non-collinear antiferromagnetic spintronics},
  journal   = {Nature Reviews Materials},
  volume    = {10},
  number    = {2},
  pages     = {109--127},
  year      = {2025},
  doi       = {10.1038/s41578-024-00706-w},
  url       = {https://doi.org/10.1038/s41578-024-00706-w},
  publisher = {Springer Nature}
}

@article{XIONG2022522,
title = {Antiferromagnetic spintronics: An overview and outlook},
journal = {Fundamental Research},
volume = {2},
number = {4},
pages = {522-534},
year = {2022},
issn = {2667-3258},
doi = {https://doi.org/10.1016/j.fmre.2022.03.016},
url = {https://www.sciencedirect.com/science/article/pii/S2667325822001443},
author = {Danrong Xiong and Yuhao Jiang and Kewen Shi and Ao Du and Yuxuan Yao and Zongxia Guo and Daoqian Zhu and Kaihua Cao and Shouzhong Peng and Wenlong Cai and Dapeng Zhu and Weisheng Zhao}}

@book{Hirohata2022Advances,
  editor    = {Atsufumi Hirohata},
  title     = {Advances in Antiferromagnetic Spintronics},
  publisher = {MDPI},
  address   = {Basel, Switzerland},
  year      = {2022},
  pages      = {82},
  isbn       = {978-3-0365-3750-4},
  note       = {Hardback ISBN: 978-3-0365-3749-8},
  doi        = {10.3390/books978-3-0365-3750-4},
  url        = {https://doi.org/10.3390/books978-3-0365-3750-4}
}

@article{DalDin2024,
  author  = {Dal Din, A. and Amin, O. J. and Wadley, P. and Edmonds, K. W.},
  title   = {Antiferromagnetic Spintronics and Beyond},
  journal = {npj Spintronics},
  volume  = {2},
  pages   = {25},
  year    = {2024},
  month   = jul,
  doi     = {10.1038/s44306-024-00029-0},
  url     = {https://doi.org/10.1038/s44306-024-00029-0},
  issn    = {2948-2119},
  publisher = {Nature Portfolio}
}

@article{Takei2014,
  title = {Superfluid spin transport through antiferromagnetic insulators},
  author = {Takei, So and Halperin, Bertrand I. and Yacoby, Amir and Tserkovnyak, Yaroslav},
  journal = {Phys. Rev. B},
  volume = {90},
  issue = {9},
  pages = {094408},
  numpages = {10},
  year = {2014},
  month = {Sep},
  publisher = {American Physical Society},
  doi = {10.1103/PhysRevB.90.094408},
  url = {https://link.aps.org/doi/10.1103/PhysRevB.90.094408}
}

@article{Cheng2014PRL,
  title = {Spin Pumping and Spin-Transfer Torques in Antiferromagnets},
  author = {Cheng, Ran and Xiao, Jiang and Niu, Qian and Brataas, Arne},
  journal = {Phys. Rev. Lett.},
  volume = {113},
  issue = {5},
  pages = {057601},
  numpages = {5},
  year = {2014},
  month = {Jul},
  publisher = {American Physical Society},
  doi = {10.1103/PhysRevLett.113.057601},
  url = {https://link.aps.org/doi/10.1103/PhysRevLett.113.057601}
}

@article{Sanchez2013PRB,
  title = {Spin pumping and inverse spin Hall effect in germanium},
  author = {Rojas-S\'anchez, J.-C. and Cubukcu, M. and Jain, A. and Vergnaud, C. and Portemont, C. and Ducruet, C. and Barski, A. and Marty, A. and Vila, L. and Attan\'e, J.-P. and Augendre, E. and Desfonds, G. and Gambarelli, S. and Jaffr\`es, H. and George, J.-M. and Jamet, M.},
  journal = {Phys. Rev. B},
  volume = {88},
  issue = {6},
  pages = {064403},
  numpages = {15},
  year = {2013},
  month = {Aug},
  publisher = {American Physical Society},
  doi = {10.1103/PhysRevB.88.064403},
  url = {https://link.aps.org/doi/10.1103/PhysRevB.88.064403}
}

@article{Sanchez2014PRL,
  title = {Spin Pumping and Inverse Spin Hall Effect in Platinum: The Essential Role of Spin-Memory Loss at Metallic Interfaces},
  author = {Rojas-S\'anchez, J.-C. and Reyren, N. and Laczkowski, P. and Savero, W. and Attan\'e, J.-P. and Deranlot, C. and Jamet, M. and George, J.-M. and Vila, L. and Jaffr\`es, H.},
  journal = {Phys. Rev. Lett.},
  volume = {112},
  issue = {10},
  pages = {106602},
  numpages = {5},
  year = {2014},
  month = {Mar},
  publisher = {American Physical Society},
  doi = {10.1103/PhysRevLett.112.106602},
  url = {https://link.aps.org/doi/10.1103/PhysRevLett.112.106602}
}

@article{Ando2011,
    author = {Ando, K. and Takahashi, S. and Ieda, J. and Kajiwara, Y. and Nakayama, H. and Yoshino, T. and Harii, K. and Fujikawa, Y. and Matsuo, M. and Maekawa, S. and Saitoh, E.},
    title = {Inverse spin-Hall effect induced by spin pumping in metallic system},
    journal = {Journal of Applied Physics},
    volume = {109},
    number = {10},
    pages = {103913},
    year = {2011},
    month = {05},
    doi = {10.1063/1.3587173},
    url = {https://doi.org/10.1063/1.3587173},
}

@book{Maekawa2012SpinCurrent,
  editor    = {Sadamichi Maekawa and Sergio O. Valenzuela and Eiji Saitoh and Takashi Kimura},
  title     = {Spin Current},
  publisher = {Oxford University Press},
  address   = {Oxford, UK},
  year      = {2012},
  isbn      = {9780199600380},
  doi       = {10.1093/acprof:oso/9780199600380.001.0001},
  url       = {https://doi.org/10.1093/acprof:oso/9780199600380.001.0001}
}

@article{Chen2015PRL,
  title = {Spin Pumping in the Presence of Spin-Orbit Coupling},
  author = {Chen, Kai and Zhang, Shufeng},
  journal = {Phys. Rev. Lett.},
  volume = {114},
  issue = {12},
  pages = {126602},
  numpages = {5},
  year = {2015},
  month = {Mar},
  publisher = {American Physical Society},
  doi = {10.1103/PhysRevLett.114.126602},
  url = {https://link.aps.org/doi/10.1103/PhysRevLett.114.126602}
}

@article{Mosendz2010,
  title = {Quantifying Spin Hall Angles from Spin Pumping: Experiments and Theory},
  author = {Mosendz, O. and Pearson, J. E. and Fradin, F. Y. and Bauer, G. E. W. and Bader, S. D. and Hoffmann, A.},
  journal = {Phys. Rev. Lett.},
  volume = {104},
  issue = {4},
  pages = {046601},
  numpages = {4},
  year = {2010},
  month = {Jan},
  publisher = {American Physical Society},
  doi = {10.1103/PhysRevLett.104.046601},
  url = {https://link.aps.org/doi/10.1103/PhysRevLett.104.046601}
}

@article{Yaroslav2002PRB,
  title = {Spin pumping and magnetization dynamics in metallic multilayers},
  author = {Tserkovnyak, Yaroslav and Brataas, Arne and Bauer, Gerrit E. W.},
  journal = {Phys. Rev. B},
  volume = {66},
  issue = {22},
  pages = {224403},
  numpages = {10},
  year = {2002},
  month = {Dec},
  publisher = {American Physical Society},
  doi = {10.1103/PhysRevB.66.224403},
  url = {https://link.aps.org/doi/10.1103/PhysRevB.66.224403}
}

@article{Kachkachi2025,
  title = {{Magnetization nutation in magnetic semiconductors: Effective spin model with anisotropic RKKY exchange interaction}},
  author = {Kachkachi, H.},
  journal = {Phys. Rev. B},
  volume = {111},
  issue = {1},
  pages = {014410},
  numpages = {17},
  year = {2025},
  month = {Jan},
  publisher = {American Physical Society},
  doi = {10.1103/PhysRevB.111.014410},
  url = {https://link.aps.org/doi/10.1103/PhysRevB.111.014410}
}

@article{Jansen2025,
  title = {Dynamically generated spin interactions and nutational spin inertia in normal metal--ferromagnet heterostructures},
  author = {Johnsen, Christian Svingen and Sudb\o{}, Asle},
  journal = {Phys. Rev. B},
  volume = {111},
  issue = {14},
  pages = {144423},
  numpages = {14},
  year = {2025},
  month = {Apr},
  publisher = {American Physical Society},
  doi = {10.1103/PhysRevB.111.144423},
  url = {https://link.aps.org/doi/10.1103/PhysRevB.111.144423}
}

@article{De2025PRB,
  title = {Magnetic nutation: Transient separation of magnetization from its angular momentum},
  author = {De, Anulekha and Schlegel, Julius and Lentfert, Akira and Scheuer, Laura and Stadtm\"uller, Benjamin and Pirro, Philipp and von Freymann, Georg and Nowak, Ulrich and Aeschlimann, Martin},
  journal = {Phys. Rev. B},
  volume = {111},
  issue = {1},
  pages = {014432},
  numpages = {7},
  year = {2025},
  month = {Jan},
  publisher = {American Physical Society},
  doi = {10.1103/PhysRevB.111.014432},
  url = {https://link.aps.org/doi/10.1103/PhysRevB.111.014432}
}

@article{Bajaj2024,
 author = {Bajaj, Robin and Lee, Seung-Cheol and Krishnamurthy, H. R. and Bhattacharjee, Satadeep and Jain, Manish},
 doi = {10.1103/PhysRevB.109.214432},
 issue = {21},
 journal = {Phys. Rev. B},
 month = {Jun},
 numpages = {10},
 pages = {214432},
 publisher = {American Physical Society},
 title = {Calculation of Gilbert damping and magnetic moment of inertia using the torque-torque correlation model within an ab initio Wannier framework},
 url = {https://link.aps.org/doi/10.1103/PhysRevB.109.214432},
 volume = {109},
 year = {2024}
}

@article{Bajpai2019,
 author = {Bajpai, Utkarsh and Nikoli\ifmmode \acute{c}\else \'{c}\fi{}, Branislav K.},
 doi = {10.1103/PhysRevB.99.134409},
 issue = {13},
 journal = {Phys. Rev. B},
 month = {Apr},
 numpages = {11},
 pages = {134409},
 publisher = {American Physical Society},
 title = {{Time-retarded damping and magnetic inertia in the {Landau-Lifshitz-Gilbert} equation self-consistently coupled to electronic time-dependent nonequilibrium Green functions}},
 url = {https://link.aps.org/doi/10.1103/PhysRevB.99.134409},
 volume = {99},
 year = {2019}
}

@article{Baltz2018,
 author = {Baltz, V. and Manchon, A. and Tsoi, M. and Moriyama, T. and Ono, T. and Tserkovnyak, Y.},
 doi = {10.1103/RevModPhys.90.015005},
 issue = {1},
 journal = {Rev. Mod. Phys.},
 month = {Feb},
 numpages = {57},
 pages = {015005},
 publisher = {American Physical Society},
 title = {Antiferromagnetic spintronics},
 url = {https://link.aps.org/doi/10.1103/RevModPhys.90.015005},
 volume = {90},
 year = {2018}
}

@article{Bhattacharjee_2012,
 author = {Bhattacharjee, Satadeep and Nordstr\"om, Lars and Fransson, Jonas},
 doi = {10.1103/PhysRevLett.108.057204},
 issue = {5},
 journal = {Phys. Rev. Lett.},
 month = {Jan},
 numpages = {5},
 pages = {057204},
 publisher = {American Physical Society},
 title = {Atomistic Spin Dynamic Method with both Damping and Moment of Inertia Effects Included from First Principles},
 url = {https://link.aps.org/doi/10.1103/PhysRevLett.108.057204},
 volume = {108},
 year = {2012}
}

@article{cherkasskii2020nutation,
 author = {Cherkasskii, Mikhail and Farle, Michael and Semisalova, Anna},
 doi = {10.1103/PhysRevB.102.184432},
 issue = {18},
 journal = {Phys. Rev. B},
 month = {Nov},
 numpages = {5},
 pages = {184432},
 publisher = {American Physical Society},
 title = {Nutation resonance in ferromagnets},
 url = {https://link.aps.org/doi/10.1103/PhysRevB.102.184432},
 volume = {102},
 year = {2020}
}

@article{Cherkasskii2022Anisotropy,
 author = {Cherkasskii, Mikhail and Barsukov, Igor and Mondal, Ritwik and Farle, Michael and Semisalova, Anna},
 doi = {10.1103/PhysRevB.106.054428},
 issue = {5},
 journal = {Phys. Rev. B},
 month = {Aug},
 numpages = {10},
 pages = {054428},
 publisher = {American Physical Society},
 title = {Theory of inertial spin dynamics in anisotropic ferromagnets},
 url = {https://link.aps.org/doi/10.1103/PhysRevB.106.054428},
 volume = {106},
 year = {2022}
}

@article{Ciornei2011,
 author = {Ciornei, M.-C. and Rub\'{\i}, J. M. and
Wegrowe, J.-E.},
 doi = {10.1103/PhysRevB.83.020410},
 journal = {Phys. Rev. B},
 month = {Jan},
 pages = {020410},
 publisher = {American Physical Society},
 title = {Magnetization dynamics in the inertial regime:
Nutation predicted at short time scales},
 url = {http://link.aps.org/doi/10.1103/PhysRevB.83.020410},
 volume = {83},
 year = {2011}
}

@article{Fahnle2011,
 author = {F\"ahnle, Manfred and Steiauf, Daniel and
Illg, Christian},
 doi = {10.1103/PhysRevB.84.172403},
 journal = {Phys. Rev. B},
 month = {Nov},
 pages = {172403},
 publisher = {American Physical Society},
 title = {{Generalized Gilbert equation including inertial
damping: Derivation from an extended breathing Fermi
surface model}},
 url = {http://link.aps.org/doi/10.1103/PhysRevB.84.172403},
 volume = {84},
 year = {2011}
}

@article{He2025,
  title = {Chirality and polarization of inertial antiferromagnetic resonances driven by spin-orbit torques},
  author = {He, Peng-Bin and Wang, Ri-Xing and Li, Zai-Dong and Cherkasskii, Mikhail},
  journal = {Phys. Rev. B},
  volume = {112},
  issue = {22},
  pages = {224423},
  numpages = {14},
  year = {2025},
  month = {Dec},
  publisher = {American Physical Society},
  doi = {10.1103/1cjr-7cgl},
  url = {https://link.aps.org/doi/10.1103/1cjr-7cgl}
}

@article{Wildes2006,
  title = {{Static and dynamic critical properties of the quasi-two-dimensional antiferromagnet ${\mathrm{MnPS}}_{3}$}},
  author = {Wildes, A. R. and R\o{}nnow, H. M. and Roessli, B. and Harris, M. J. and Godfrey, K. W.},
  journal = {Phys. Rev. B},
  volume = {74},
  issue = {9},
  pages = {094422},
  numpages = {13},
  year = {2006},
  month = {Sep},
  publisher = {American Physical Society},
  doi = {10.1103/PhysRevB.74.094422},
  url = {https://link.aps.org/doi/10.1103/PhysRevB.74.094422}
}

@article{Kim_2019,
doi = {10.1088/2053-1583/ab27d5},
url = {https://doi.org/10.1088/2053-1583/ab27d5},
year = {2019},
month = {jul},
publisher = {IOP Publishing},
volume = {6},
number = {4},
pages = {041001},
author = {Kim, Kangwon and Lim, Soo Yeon and Kim, Jungcheol and Lee, Jae-Ung and Lee, Sungmin and Kim, Pilkwang and Park, Kisoo and Son, Suhan and Park, Cheol-Hwan and Park, Je-Geun and Cheong, Hyeonsik},
title = {{Antiferromagnetic ordering in van der Waals 2D magnetic material MnPS$_3$ probed by Raman spectroscopy}},
journal = {2D Materials}
}

@article{Johansen2017,
  title = {Spin pumping and inverse spin Hall voltages from dynamical antiferromagnets},
  author = {Johansen, \O{}yvind and Brataas, Arne},
  journal = {Phys. Rev. B},
  volume = {95},
  issue = {22},
  pages = {220408(R)},
  numpages = {5},
  year = {2017},
  month = {Jun},
  publisher = {American Physical Society},
  doi = {10.1103/PhysRevB.95.220408},
  url = {https://link.aps.org/doi/10.1103/PhysRevB.95.220408}
}

@article{Alliati2022MnPS3,
  author  = {Alliati, Ignacio M. and Evans, Richard F. L. and Novoselov, Kostya S. and Santos, Elton J. G.},
  title   = {Relativistic domain-wall dynamics in van der Waals antiferromagnet {MnPS$_3$}},
  journal = {npj Computational Materials},
  volume  = {8},
  number  = {1},
  pages   = {3},
  year    = {2022},
  doi     = {10.1038/s41524-021-00683-6}
}

@article{Varela-Manjarres_2023,
doi = {10.1088/2515-7639/aceaad},
url = {https://doi.org/10.1088/2515-7639/aceaad},
year = {2023},
month = {aug},
publisher = {IOP Publishing},
volume = {6},
number = {4},
pages = {045001},
author = {Varela-Manjarres, Jalil and Nikolić, Branislav K},
title = {High-harmonic generation in spin and charge current pumping at ferromagnetic or antiferromagnetic resonance in the presence of spin–orbit coupling},
journal = {Journal of Physics: Materials}
}

@article{Ominato_2025,
doi = {10.1088/1361-648X/ae1090},
url = {https://doi.org/10.1088/1361-648X/ae1090},
year = {2025},
month = {oct},
publisher = {IOP Publishing},
volume = {37},
number = {43},
pages = {433001},
author = {Ominato, Yuya and Yama, Masaki and Yamakage, Ai and Matsuo, Mamoru and Kato, Takeo},
title = {Spin pumping into two-dimensional systems},
journal = {Journal of Physics: Condensed Matter}
}

@article{Chen2021MagnonValve,
  author  = {Chen, Guangyi and Qi, Shaomian and Liu, Jianqiao and Chen, Di
             and Wang, Jiongjie and Yan, Shili and Zhang, Yu and Cao, Shimin
             and Lu, Ming and Tian, Shibing and Chen, Kangyao and Yu, Peng
             and Liu, Zheng and Xie, X. C. and Xiao, Jiang and Shindou, Ryuichi
             and Chen, Jian-Hao},
  title   = {Electrically switchable van der Waals magnon valves},
  journal = {Nature Communications},
  volume  = {12},
  pages   = {6279},
  year    = {2021},
  doi     = {10.1038/s41467-021-26523-1}
}

@article{Kholid2023Interface,
  author  = {Kholid, Farhan Nur and Hamara, Dominik and Hamdan, Ahmad Faisal Bin and Antonio, Guillermo Nava and Bowen, Richard and Petit, Doroth{\'e}e and Cowburn, Russell and Pisarev, Roman V. and Bossini, Davide and Barker, Joseph and Ciccarelli, Chiara},
  title   = {The importance of the interface for picosecond spin pumping in antiferromagnet-heavy metal heterostructures},
  journal = {Nature Communications},
  volume  = {14},
  pages   = {538},
  year    = {2023},
  doi     = {10.1038/s41467-023-36166-z},
  url     = {https://doi.org/10.1038/s41467-023-36166-z}
}

@article{Vaidya2020Subterahertz,
  author  = {Vaidya, Priyanka and Morley, Sophie A. and van Tol, Johan and Liu, Yan and Cheng, Ran and Brataas, Arne and Lederman, David and del Barco, Enrique},
  title   = {Subterahertz Spin Pumping from an Insulating Antiferromagnet},
  journal = {Science},
  volume  = {368},
  number  = {6487},
  pages   = {160--165},
  year    = {2020},
  doi     = {10.1126/science.aaz4247},
  url     = {https://doi.org/10.1126/science.aaz4247}
}

@article{Wang2021,
  title = {Spin Pumping of an Easy-Plane Antiferromagnet Enhanced by Dzyaloshinskii--Moriya Interaction},
  author = {Wang, Hailong and Xiao, Yuxuan and Guo, Mingda and Lee-Wong, Eric and Yan, Gerald Q. and Cheng, Ran and Du, Chunhui Rita},
  journal = {Phys. Rev. Lett.},
  volume = {127},
  issue = {11},
  pages = {117202},
  numpages = {7},
  year = {2021},
  month = {Sep},
  publisher = {American Physical Society},
  doi = {10.1103/PhysRevLett.127.117202},
  url = {https://link.aps.org/doi/10.1103/PhysRevLett.127.117202}
}

@article{Subedi2025,
  title = {Engineering antiferromagnetic magnon bands through interlayer spin pumping},
  author = {Subedi, M.M. and Deng, K. and Xiong, Y. and Mongeon, J. and Hossain, M.T. and Meisenheimer, P.B. and Zhou, E.T. and Heron, J.T. and Jungfleisch, M.B. and Zhang, W. and Flebus, B. and Sklenar, J.},
  journal = {Phys. Rev. Appl.},
  volume = {23},
  issue = {3},
  pages = {L031003},
  numpages = {6},
  year = {2025},
  month = {Mar},
  publisher = {American Physical Society},
  doi = {10.1103/PhysRevApplied.23.L031003},
  url = {https://link.aps.org/doi/10.1103/PhysRevApplied.23.L031003}
}

@article{Hartmann2025NonMarkovian,
  author        = {Hartmann, A. and others},
  title         = {Intrinsic non-Markovian magnetisation dynamics},
  journal       = {arXiv preprint arXiv:2512.07378},
  year          = {2025},
  eprint        = {2512.07378},
  archivePrefix = {arXiv},
  primaryClass  = {cond-mat.mtrl-sci},
  url           = {https://arxiv.org/abs/2512.07378}
}

@article{Wildes1998SpinWaves,
  author    = {Wildes, A. R. and Roessli, B. and Lebech, B. and Godfrey, K. W.},
  title     = {Spin waves and the critical behaviour of the magnetization in {MnPS}$_3$},
  journal   = {Journal of Physics: Condensed Matter},
  volume    = {10},
  number    = {28},
  pages     = {6417--6428},
  year      = {1998},
  doi       = {10.1088/0953-8984/10/28/014},
  publisher = {IOP Publishing}
}

@article{Ghosh2024,
 author = {Ghosh, Subhadip and Cherkasskii, Mikhail and Barsukov, Igor and Mondal, Ritwik},
 doi = {10.1103/PhysRevB.110.174430},
 issue = {17},
 journal = {Phys. Rev. B},
 month = {Nov},
 numpages = {12},
 pages = {174430},
 publisher = {American Physical Society},
 title = {Theory of tensorial magnetic inertia in terahertz spin dynamics},
 url = {https://link.aps.org/doi/10.1103/PhysRevB.110.174430},
 volume = {110},
 year = {2024}
}

@article{He2023,
 author = {He, Peng-Bin},
 doi = {10.1103/PhysRevB.108.184418},
 issue = {18},
 journal = {Phys. Rev. B},
 month = {Nov},
 numpages = {11},
 pages = {184418},
 publisher = {American Physical Society},
 title = {Large-amplitude and widely tunable self-oscillations enabled by the inertial effect in uniaxial antiferromagnets driven by spin-orbit torques},
 url = {https://link.aps.org/doi/10.1103/PhysRevB.108.184418},
 volume = {108},
 year = {2023}
}

@article{He2024PRB,
 author = {He, Peng-Bin},
 doi = {10.1103/PhysRevB.110.064411},
 issue = {6},
 journal = {Phys. Rev. B},
 month = {Aug},
 numpages = {14},
 pages = {064411},
 publisher = {American Physical Society},
 title = {Influence of the magnetic inertia on the self-oscillation in spin-orbit torque-driven tripartite antiferromagnets with a ${120}^{\ensuremath{\circ}}$ rotation symmetry},
 url = {https://link.aps.org/doi/10.1103/PhysRevB.110.064411},
 volume = {110},
 year = {2024}
}

@article{Jungwirth2016,
 author = {Jungwirth, T.
and Marti, X.
and Wadley, P.
and Wunderlich, J.},
 day = {01},
 doi = {10.1038/nnano.2016.18},
 issn = {1748-3395},
 journal = {Nat. Nanotechnol.},
 month = {Mar},
 number = {3},
 pages = {231-241},
 title = {Antiferromagnetic spintronics},
 url = {https://doi.org/10.1038/nnano.2016.18},
 volume = {11},
 year = {2016}
}

@article{Kampfrath2011,
 author = {Tobias Kampfrath and Alexander Sell and Gregor Klatt and Alexej Pashkin and Sebastian M{\"a}hrlein and Thomas Dekorsy and Martin Wolf and Manfred Fiebig and Alfred Leitenstorfer and Rupert Huber},
 day = {21},
 doi = {10.1038/nphoton.2010.259},
 journal = {Nat. Photon.},
 month = {November},
 pages = {31},
 title = {Coherent terahertz control of antiferromagnetic spin waves},
 url = {https://www.nature.com/articles/nphoton.2010.259},
 volume = {5},
 year = {2011}
}

@article{Kikuchi2015,
 author = {Kikuchi, Toru and Tatara, Gen},
 doi = {10.1103/PhysRevB.92.184410},
 issue = {18},
 journal = {Phys. Rev. B},
 month = {Nov},
 numpages = {15},
 pages = {184410},
 publisher = {American Physical Society},
 title = {Spin dynamics with inertia in metallic ferromagnets},
 url = {https://link.aps.org/doi/10.1103/PhysRevB.92.184410},
 volume = {92},
 year = {2015}
}

@article{Mondal2017Nutation,
 author = {Mondal, Ritwik and Berritta, Marco and Nandy, Ashis K. and Oppeneer, Peter M.},
 doi = {10.1103/PhysRevB.96.024425},
 issue = {2},
 journal = {Phys. Rev. B},
 month = {Jul},
 numpages = {9},
 pages = {024425},
 publisher = {American Physical Society},
 title = {Relativistic theory of magnetic inertia in ultrafast spin dynamics},
 url = {https://link.aps.org/doi/10.1103/PhysRevB.96.024425},
 volume = {96},
 year = {2017}
}

@article{Mondal2018,
 abstractnote = {The phenomenological Landau-Lifshitz-Gilbert (LLG) equation of motion remains as the cornerstone of contemporary magnetisation dynamics studies, wherein the Gilbert damping parameter has been attributed to first-order relativistic effects. To include magnetic inertial effects the LLG equation has previously been extended with a supplemental inertia term; the arising inertial dynamics has been related to second-order relativistic effects. Here we start from the relativistic Dirac equation and, performing a Foldy-Wouthuysen transformation, derive a generalised Pauli spin Hamiltonian that contains relativistic correction terms to any higher order. Using the Heisenberg equation of spin motion we derive general relativistic expressions for the tensorial Gilbert damping and magnetic inertia parameters, and show that these tensors can be expressed as series of higher-order relativistic correction terms. We further show that, in the case of a harmonic external driving field, these series can be summed and we provide closed analytical expressions for the Gilbert and inertial parameters that are functions of the frequency of the driving field.},
 author = {Mondal, Ritwik and Berritta, Marco and Oppeneer, Peter M.},
 doi = {10.1088/1361-648X/aac5a2},
 issn = {0953-8984},
 journal = {Journal of Physics: Condensed Matter},
 month = {Jul},
 number = {26},
 pages = {265801},
 publisher = {IOP Publishing},
 title = {{Generalisation of {G}ilbert damping and magnetic inertia parameter as a series of higher-order relativistic terms}},
 url = {https://dx.doi.org/10.1088/1361-648X/aac5a2},
 volume = {30},
 year = {2018}
}

@article{Mondal2020nutation,
 author = {Mondal, Ritwik and Gro\ss{}enbach, Sebastian and R\'ozsa, Levente and Nowak, Ulrich},
 doi = {10.1103/PhysRevB.103.104404},
 issue = {10},
 journal = {Phys. Rev. B},
 month = {Mar},
 numpages = {11},
 pages = {104404},
 publisher = {American Physical Society},
 title = {Nutation in antiferromagnetic resonance},
 url = {https://link.aps.org/doi/10.1103/PhysRevB.103.104404},
 volume = {103},
 year = {2021}
}

@article{Mondal2021PRB,
 author = {Mondal, Ritwik and Oppeneer, Peter M.},
 doi = {10.1103/PhysRevB.104.104405},
 issue = {10},
 journal = {Phys. Rev. B},
 month = {Sep},
 numpages = {8},
 pages = {104405},
 publisher = {American Physical Society},
 title = {Influence of intersublattice coupling on the terahertz nutation spin dynamics in antiferromagnets},
 url = {https://link.aps.org/doi/10.1103/PhysRevB.104.104405},
 volume = {104},
 year = {2021}
}

@article{Mondal2021PRBSpinCurrent,
 author = {Mondal, Ritwik and Kamra, Akashdeep},
 doi = {10.1103/PhysRevB.104.214426},
 issue = {21},
 journal = {Phys. Rev. B},
 month = {Dec},
 numpages = {7},
 pages = {214426},
 publisher = {American Physical Society},
 title = {Spin pumping at terahertz nutation resonances},
 url = {https://link.aps.org/doi/10.1103/PhysRevB.104.214426},
 volume = {104},
 year = {2021}
}

@article{Mondal2021JPCM,
 author = {Mondal, Ritwik},
 doi = {10.1088/1361-648X/abfc6d},
 journal = {J. Phys.: Condens. Matter},
 month = {may},
 number = {27},
 pages = {275804},
 publisher = {IOP Publishing},
 title = {Theroy of magnetic inertial dynamics in two-sublattice ferromagnets},
 url = {https://dx.doi.org/10.1088/1361-648X/abfc6d},
 volume = {33},
 year = {2021}
}

@article{MONDAL_Review,
 author = {Ritwik Mondal and Levente Rózsa and Michael Farle and Peter M. Oppeneer and Ulrich Nowak and Mikhail Cherkasskii},
 doi = {https://doi.org/10.1016/j.jmmm.2023.170830},
 issn = {0304-8853},
 journal = {J. Magn. Magn. Mater.},
 pages = {170830},
 title = {Inertial effects in ultrafast spin dynamics},
 url = {https://www.sciencedirect.com/science/article/pii/S0304885323004791},
 volume = {579},
 year = {2023}
}

@article{Neeraj2020,
 abstractnote = {The understanding of how spins move and can be manipulated at pico- and femtosecond timescales has implications for ultrafast and energy-efficient data-processing and storage applications. However, the possibility of realizing commercial technologies based on ultrafast spin dynamics has been hampered by our limited knowledge of the physics behind processes on this timescale. Recently, it has been suggested that inertial effects should be considered in the full description of the spin dynamics at these ultrafast timescales, but a clear observation of such effects in ferromagnets is still lacking. Here, we report direct experimental evidence of intrinsic inertial spin dynamics in ferromagnetic thin films in the form of a nutation of the magnetization at a frequency of ~0.5 THz. This allows us to reveal that the angular momentum relaxation time in ferromagnets is on the order of 10 ps.},
 author = {Neeraj, Kumar and Awari, Nilesh and Kovalev, Sergey and Polley, Debanjan and Zhou Hagström, Nanna and Arekapudi, Sri Sai Phani Kanth and Semisalova, Anna and Lenz, Kilian and Green, Bertram and Deinert, Jan Christoph and et al.},
 doi = {10.1038/s41567-020-01040-y},
 issn = {17452481},
 journal = {Nature Physics},
 number = {2},
 pages = {245–250},
 publisher = {Nature Research},
 title = {Inertial spin dynamics in ferromagnets},
 volume = {17},
 year = {2020}
}

@article{Olive2012,
 author = {Olive,E.  and Lansac,Y.  and Wegrowe,J.-E. },
 doi = {10.1063/1.4712056},
 journal = {Appl. Phys. Lett.},
 number = {19},
 pages = {192407},
 title = {Beyond ferromagnetic resonance: The inertial regime of the magnetization},
 url = {https://doi.org/10.1063/1.4712056},
 volume = {100},
 year = {2012}
}

@article{Olive2015,
 author = {Olive,E.  and Lansac,Y.  and Meyer,M.  and Hayoun,M.  and Wegrowe,J.-E. },
 doi = {10.1063/1.4921908},
 journal = {J. Appl. Phys.},
 number = {21},
 pages = {213904},
 title = {{Deviation from the Landau-Lifshitz-Gilbert equation in the inertial regime of the magnetization}},
 url = { 
https://doi.org/10.1063/1.4921908
},
 volume = {117},
 year = {2015}
}

@article{Mondal2022PRB,
 author = {Mondal, Ritwik and R\'ozsa, Levente},
 doi = {10.1103/PhysRevB.106.134422},
 issue = {13},
 journal = {Phys. Rev. B},
 month = {Oct},
 numpages = {14},
 pages = {134422},
 publisher = {American Physical Society},
 title = {Inertial spin waves in ferromagnets and antiferromagnets},
 url = {https://link.aps.org/doi/10.1103/PhysRevB.106.134422},
 volume = {106},
 year = {2022}
}

@article{Rodriguez2024PRL,
 author = {Rodriguez, Rodolfo and Cherkasskii, Mikhail and Jiang, Rundong and Mondal, Ritwik and Etesamirad, Arezoo and Tossounian, Allison and Ivanov, Boris A. and Barsukov, Igor},
 doi = {10.1103/PhysRevLett.132.246701},
 issue = {24},
 journal = {Phys. Rev. Lett.},
 month = {Jun},
 numpages = {6},
 pages = {246701},
 publisher = {American Physical Society},
 title = {Spin Inertia and Auto-Oscillations in Ferromagnets},
 url = {https://link.aps.org/doi/10.1103/PhysRevLett.132.246701},
 volume = {132},
 year = {2024}
}

@article{Rozsa_2013,
 author = {Rózsa, L and Udvardi, L and Szunyogh, L},
 doi = {10.1088/0953-8984/25/50/506002},
 journal = {J. Phys.: Condens. Matter},
 month = {nov},
 number = {50},
 pages = {506002},
 publisher = {IOP Publishing},
 title = {Relativistic and thermal effects on the magnon spectrum of a ferromagnetic monolayer},
 url = {https://dx.doi.org/10.1088/0953-8984/25/50/506002},
 volume = {25},
 year = {2013}
}

@article{Kajiwara2010,
  author  = {Kajiwara, Y. and Harii, K. and Takahashi, S. and Ohe, J. and
             Uchida, K. and Mizuguchi, M. and Umezawa, H. and Kawai, H. and
             Ando, K. and Takanashi, K. and Maekawa, S. and Saitoh, E.},
  title   = {Transmission of electrical signals by spin-wave interconversion
             in a magnetic insulator},
  journal = {Nature},
  volume  = {464},
  pages   = {262--266},
  year    = {2010},
  doi     = {10.1038/nature08876},
}

@article{Heinrich2011,
  title = {Spin Pumping at the Magnetic Insulator (YIG)/Normal Metal (Au) Interfaces},
  author = {Heinrich, B. and Burrowes, C. and Montoya, E. and Kardasz, B. and Girt, E. and Song, Young-Yeal and Sun, Yiyan and Wu, Mingzhong},
  journal = {Phys. Rev. Lett.},
  volume = {107},
  issue = {6},
  pages = {066604},
  numpages = {4},
  year = {2011},
  month = {Aug},
  publisher = {American Physical Society},
  doi = {10.1103/PhysRevLett.107.066604},
  url = {https://link.aps.org/doi/10.1103/PhysRevLett.107.066604}
}

@article{Cho2023,
  title = {Spin pumping from a ferromagnetic insulator into an altermagnet},
  author = {Sun, Chi and Linder, Jacob},
  journal = {Phys. Rev. B},
  volume = {108},
  issue = {14},
  pages = {L140408},
  numpages = {5},
  year = {2023},
  month = {Oct},
  publisher = {American Physical Society},
  doi = {10.1103/PhysRevB.108.L140408},
  url = {https://link.aps.org/doi/10.1103/PhysRevB.108.L140408}
}

@article{Hodt2024,
  title = {Spin pumping in an altermagnet/normal-metal bilayer},
  author = {Hodt, Erik Wegner and Linder, Jacob},
  journal = {Phys. Rev. B},
  volume = {109},
  issue = {17},
  pages = {174438},
  numpages = {13},
  year = {2024},
  month = {May},
  publisher = {American Physical Society},
  doi = {10.1103/PhysRevB.109.174438},
  url = {https://link.aps.org/doi/10.1103/PhysRevB.109.174438}
}

@article{Ghosh2026PRR,
  title = {Spin inertia as a source of topological magnons: Chiral edge states from coupled precession and nutation},
  author = {Ghosh, Subhadip and Cherkasskii, Mikhail and Mondal, Ritwik and Mook, Alexander and R\'ozsa, Levente},
  journal = {Phys. Rev. Res.},
  volume = {8},
  issue = {3},
  pages = {L032040},
  numpages = {6},
  year = {2026},
  month = {Sep},
  publisher = {American Physical Society},
  doi = {10.1103/7njs-vyl9},
  url = {https://link.aps.org/doi/10.1103/7njs-vyl9}
}

@article{Gupta2024,
  title = {{Self-induced inverse spin Hall effect in ${\mathrm{La}}_{0.67}{\mathrm{Sr}}_{0.33}{\mathrm{MnO}}_{3}$ films}},
  author = {Gupta, Pushpendra and Park, In Jun and Swain, Anupama and Mishra, Abhisek and Amin, Vivek P. and Bedanta, Subhankar},
  journal = {Phys. Rev. B},
  volume = {109},
  issue = {1},
  pages = {014437},
  numpages = {7},
  year = {2024},
  month = {Jan},
  publisher = {American Physical Society},
  doi = {10.1103/PhysRevB.109.014437},
  url = {https://link.aps.org/doi/10.1103/PhysRevB.109.014437}
}

@article{Thibaudeau2021,
 author = {Thibaudeau, Pascal and Nicolis, Stam},
 doi = {10.1140/epjb/s10051-021-00211-x},
 journal = {The European Physical Journal B},
 month = {oct},
 number = {10},
 pages = {196},
 title = {{Emerging magnetic nutation}},
 url = {https://link.springer.com/10.1140/epjb/s10051-021-00211-x},
 volume = {94},
 year = {2021}
}

@article{Yang_2017,
doi = {10.1088/1367-2630/aa5487},
url = {https://doi.org/10.1088/1367-2630/aa5487},
year = {2017},
month = {jan},
publisher = {IOP Publishing},
volume = {19},
number = {1},
pages = {015008},
author = {Yang, Hao and Sun, Yan and Zhang, Yang and Shi, Wu-Jun and Parkin, Stuart S P and Yan, Binghai},
title = {{Topological Weyl semimetals in the chiral antiferromagnetic materials Mn$_3$Ge and Mn$_3$Sn}},
journal = {New Journal of Physics}
}

@article{Libor2020,
author = {Libor Šmejkal  and Rafael González-Hernández  and T. Jungwirth  and J. Sinova },
title = {Crystal time-reversal symmetry breaking and spontaneous Hall effect in collinear antiferromagnets},
journal = {Science Advances},
volume = {6},
number = {23},
pages = {eaaz8809},
year = {2020},
doi = {10.1126/sciadv.aaz8809},
URL = {https://www.science.org/doi/abs/10.1126/sciadv.aaz8809}}

@article{thonig2017magnetic,
 author = {Thonig, Danny and Eriksson, Olle and Pereiro, Manuel},
 journal = {Scientific reports},
 number = {1},
 pages = {931},
 publisher = {Nature Publishing Group UK London},
 title = {Magnetic moment of inertia within the torque-torque correlation model},
 url = {https://doi.org/10.1038/s41598-017-01081-z},
 volume = {7},
 year = {2017}
}

@article{Titov2024JAP,
 author = {Titov, Sergei V. and Dowling, William J. and Titov, Anton S. and Fedorov, Andrey S.},
 doi = {10.1063/5.0196622},
 issn = {0021-8979},
 journal = {J. Appl. Phys.},
 month = {03},
 number = {9},
 pages = {093903},
 title = {{Antiferromagnetic and nutation resonance frequencies of antiferromagnets at an arbitrary strength of the applied dc field}},
 url = {https://doi.org/10.1063/5.0196622},
 volume = {135},
 year = {2024}
}

@article{Titov_2022,
 author = {Titov, Sergei V. and Dowling, William J. and Kalmykov, Yuri P. and Cherkasskii, Mikhail},
 doi = {10.1103/PhysRevB.105.214414},
 issue = {21},
 journal = {Phys. Rev. B},
 month = {Jun},
 numpages = {9},
 pages = {214414},
 publisher = {American Physical Society},
 title = {Nutation spin waves in ferromagnets},
 url = {https://link.aps.org/doi/10.1103/PhysRevB.105.214414},
 volume = {105},
 year = {2022}
}

@article{Tserkovnyak2005,
 author = {Tserkovnyak, Yaroslav and Brataas, Arne and Bauer, Gerrit E. W. and Halperin, Bertrand I.},
 doi = {10.1103/RevModPhys.77.1375},
 issue = {4},
 journal = {Rev. Mod. Phys.},
 month = {Dec},
 numpages = {0},
 pages = {1375--1421},
 publisher = {American Physical Society},
 title = {Nonlocal magnetization dynamics in ferromagnetic heterostructures},
 url = {https://link.aps.org/doi/10.1103/RevModPhys.77.1375},
 volume = {77},
 year = {2005}
}

@article{unikandanunni2021inertial,
 author = {Unikandanunni, Vivek and Medapalli, Rajasekhar and Asa, Marco and Albisetti, Edoardo and Petti, Daniela and Bertacco, Riccardo and Fullerton, Eric E. and Bonetti, Stefano},
 doi = {10.1103/PhysRevLett.129.237201},
 issue = {23},
 journal = {Phys. Rev. Lett.},
 month = {Nov},
 numpages = {6},
 pages = {237201},
 publisher = {American Physical Society},
 title = {Inertial Spin Dynamics in Epitaxial Cobalt Films},
 url = {https://link.aps.org/doi/10.1103/PhysRevLett.129.237201},
 volume = {129},
 year = {2022}
}

@article{Winter2022,
 author = {Winter, Lucas and Gro\ss{}enbach, Sebastian and Nowak, Ulrich and R\'ozsa, Levente},
 doi = {10.1103/PhysRevB.106.214403},
 issue = {21},
 journal = {Phys. Rev. B},
 month = {Dec},
 numpages = {12},
 pages = {214403},
 publisher = {American Physical Society},
 title = {Nutational switching in ferromagnets and antiferromagnets},
 url = {https://link.aps.org/doi/10.1103/PhysRevB.106.214403},
 volume = {106},
 year = {2022}
}

@article{Yan_2024,
doi = {10.1088/1361-648X/ad06ef},
url = {https://doi.org/10.1088/1361-648X/ad06ef},
year = {2023},
month = {nov},
publisher = {IOP Publishing},
volume = {36},
number = {6},
pages = {065502},
author = {Yan, Songsong and Du, Yongping and Zhang, Xiaoou and Wan, Xiangang and Wang, Di},
title ={ {First-principles study of magnetic interactions and excitations in antiferromagnetic van der Waals material MPX3 (M=Mn, Fe, Co, Ni; X=S, Se)}},
journal = {Journal of Physics: Condensed Matter}
}

@article{Wang2023,
  title = {{Exciton-magnon splitting in the van der Waals antiferromagnet ${\mathrm{MnPS}}_{3}$ unveiled by second-harmonic generation}},
  author = {Wang, Ziqian and Zhang, Xiao-Xiao and Shiomi, Yuki and Arima, Taka-hisa and Nagaosa, Naoto and Tokura, Yoshinori and Ogawa, Naoki},
  journal = {Phys. Rev. Res.},
  volume = {5},
  issue = {4},
  pages = {L042032},
  numpages = {6},
  year = {2023},
  month = {Nov},
  publisher = {American Physical Society},
  doi = {10.1103/PhysRevResearch.5.L042032},
  url = {https://link.aps.org/doi/10.1103/PhysRevResearch.5.L042032}
}

@article{BABUKA2020109592,
title = {{Electronic and vibrational properties of pure MnPS$_3$ crystal: Theoretical and experimental investigation}},
journal = {Computational Materials Science},
volume = {177},
pages = {109592},
year = {2020},
issn = {0927-0256},
doi = {https://doi.org/10.1016/j.commatsci.2020.109592},
url = {https://www.sciencedirect.com/science/article/pii/S0927025620300835},
author = {T. Babuka and M. Makowska-Janusik and A.V. Peschanskii and K.E. Glukhov and S.L. Gnatchenko and Yu.M. Vysochanskii}
}
\end{document}